\documentclass[twocolumn,showpacs,pra]{revtex4}

\begin{document}

\newcommand{\ket}[1]{\ensuremath{|#1 \rangle}}
\newcommand{\bra}[1]{\ensuremath{\langle #1|}}
\newcommand{\braket}[2]{\ensuremath{\langle #1|#2 \rangle}}
\newcommand{\ketbra}[2]{\ensuremath{|#1 \rangle \langle #2|}}
\newcommand{\ro}[1]{\ensuremath{|#1 \rangle \langle #1|}}
\newcommand{\av}[1]{\ensuremath{\langle #1 \rangle}}
\newcommand{\trace}{\ensuremath{\textsf{Tr}}}
\newcommand{\mi}{\ensuremath{\mathrm{i}}}
\newcommand{\q}{\ensuremath{q_{ai}}}
\newcommand{\dd}[2]{\ensuremath{\frac{\mathrm{d}#1}{\mathrm{d}#2}}}
\newcommand{\R}{\ensuremath{{\sf R\hspace*{-0.9ex}\rule{0.15ex}
{1.5ex}\hspace*{0.9ex}}}}
\newcommand{\h}[1]{\ensuremath{\mathcal{H}_{#1}}}
\newcommand{\twtw}[4]{\ensuremath{\left( \begin{array}{cc} #1 & #2 \\
#3 & #4 \end{array} \right)}}

\title{Developing the Deutsch-Hayden approach to quantum mechanics}
\author{C. Hewitt-Horsman$^{\dagger}$, V. Vedral$^{\diamond}$}
\address{$^{\dagger}$Blackett Laboratory,
Imperial College, London SW7 2BZ, United Kingdom\\
$^{\diamond}$School of Physics and Astronomy, University of Leeds, Leeds LS2 9JT, United Kingdom}
\date{13 September 2006}

\begin{abstract} The formalism of Deutsch and Hayden is a useful tool for describing quantum mechanics explicitly as local and unitary, and therefore quantum information theory as concerning a ``flow" of information between systems. In this paper we show that these physical descriptions of flow are unique, and develop the approach further to include the measurement interaction and mixed states. We then give an analysis of entanglement swapping in this approach, showing that it does not in fact contain non-local effects or some form of superluminal signalling.

\end{abstract}

\maketitle

\section{Introduction}

The Deutsch-Hayden approach \cite{dh} was introduced in order to describe the flow of information through quantum mechanical systems. The standard formalism is not suitable in this situation as it has non-local states; that is, a single state can describe systems which are physically far apart. Any information stored in that state can then appear to `jump' from one system to the other without an explicit physical communication happening. Deutsch and Hayden demonstrated that this does not happen in their approach, and that only local interactions change the description of a system. Furthermore, unlike in the standard formalism, there is no notion of ``collapse'' in the Deutsch-Hayden approach. All evolution is unitary, even under the action of measurement. This approach is therefore very useful if we wish a formalism of quantum mechanics that is explicitly local and unitary. One particular use of such a formalism is in information theory. As Deutsch and Hayden showed in their analysis of teleportation, if we have no notion of collapse then `bits' and `classical communication' are treated by the formalism in exactly the same way as qubits and communication via quantum channels. This is very useful as we no longer need to swap between different types of entities in the middle of a protocol.

Deutsch and Hayden demonstrated information flow in some situations, for example in teleportation. However, there were some situations that they did not address, and some issues that need to be resolved. Firstly, the demonstration of information flow depends on the exact form of the description of the systems. However, if we are to use this then we need to know how unique those forms are: if another form may equally well be used, where does that leave the analysis? Then there are issues of measurement. The measurement interaction itself is unitary, but eventually in order to extract prediction from the theory we are going to need something that will stand in for collapse -- a notion of what the description of one system is relative to a description of the other. Finally, Deutsch and Hayden dealt only with pure state systems, so in order to use the formalism universally we are going to need a way of describing mixed state systems within the approach.

These, then, are the issues that this paper will address. We will first give an introduction to Deutsch and Hayden's work. We will then look at the uniqueness of the descriptions given. Then we will deal with the measurement interaction and relative states. After this, we will look in detail at how mixed state systems may be incorporated into the formalism. We will then use the developed formalism to analyse information flow in a protocol that could not be described fully previously: that of entanglement swapping.

\section{The Deutsch-Hayden approach}

The Deutsch-Hayden approach is based on the formalism first introduced by Gottesman \cite{gottesman} in the context of stabiliser theory, and is a Heisenberg-type representation of quantum mechanics. Instead of the states of systems evolving under a Hamiltonian and operators being fixed, there is a fixed universal state (by convention usually $\ket{\mathbf{0}}$), and an operator $A(t)$ evolves over time with equations of motion
$$ \mi \hbar \dd{A(t)}{t} = [A(t),H]$$

\noindent where $H$ is the Hamiltonian. We can also write the time-dependent operator as $A(t) = U^\dagger A(0) U$, where $U$ is the unitary evolution operator.

The Deutsch-Hayden representation in based on the Hilbert space where these time-dependent operators are vectors rather than second rank tensors -- the space of Hilbert-Schmidt operators \cite{fano}. This Hilbert-Schmidt space has an inner product ($\trace (A^\dagger B)$ where $A$ and $B$ are operators) and a norm ($\sqrt{\trace(A^2)}$), and for an $N$-dimensional system an $N^2$ dimensional space is required. 

As an example, let us consider a single qubit. Such a system is 2 dimensional and will therefore need a set of 4 basis operators. One such set that would be useful is that of the 2 dimensional Pauli operators, $\{ \openone , \mathbf{\sigma} \}$. Now, the $\openone$ component of any operator can never change (it will evolve as $U^\dagger \openone U = \openone$), and in order for the operator to be normalised, the $\openone$ component must always be $1/N^2$. Thus a qubit can be characterised by giving the components in the $\sigma_x, \sigma_y, \sigma_z$ directions only. This is, of course, the Bloch sphere \cite[p174]{nandc}.

In the general case, we choose a set of basis operators written here as $\Gamma_i$. Then any operator can be written in terms of the basis (here at time $t=0$):
$$ A(0) = \sum_{i=0}^{N^2-1} a_i \Gamma_i (0)$$

\noindent where $a_i \in \R$ is the inner product $\trace (A(0)\Gamma_i(0))$ (we assume that all the Hilbert-Schmidt operators with which we work are self-adjoint; that is, they may be the operators corresponding to observables). If we now evolve $A(0)$ to time $t$ we have
$$A(t) = U^\dagger A(0) U  = \sum_{i=0}^{N^2-1} a_i \Gamma_i(t) $$

\noindent Thus we see that in order to find the time-evolved state of an operator, all we need do is find the time-evolved basis operators and then reconstruct the operator using the original coefficients. Thus a complete characterisation of the time-evolved system can be gained by following the time evolution of the basis operators.

However, this is not usually very practical: as $N$ increases, the task of following the evolution of $N^2$ basis operators quickly becomes unwieldy. It was Gottesman \cite{gottesman} who realised that the number of operators to be tracked can be reduced further. Operators can always be written as combinations of other operators, either additively or multipicatively. It is obvious that for additive groups of operators the evolution operator preserves the group structure (\emph{ie} $U^\dagger(A+B)U = U^\dagger A U + U^\dagger B U$). However it is also true of the multiplicative groups, where $U$ forms a group homomorphism between $\{X\}$ and $\{ U^\dagger X U\}$:
\begin{eqnarray*} A(0)B(0) \rightarrow U^\dagger AB U & = &  U^\dagger A UU^\dagger B U \\
{} & = & A(t)B(t) \end{eqnarray*}

We can therefore see that, rather than following the complete set of basis operators, all we need follow is a generating set of the group. The time-evolved generating set will preserve the group structure of the complete set, enabling that to be constructed out of it.The generating set that Gottesman (whom Deutsch and Hayden follow) chooses is the Pauli group. This gives us the representation of an $n$-qubit array, each qubit of which is defined by a set of operators that will generate all the operators pertaining to that qubit. We note that this representation does in fact have a further redundancy within itself, as the 2-dimensional Pauli operators themselves form a multiplicative group (\emph{eg} $\sigma_y = \mi \sigma_x \sigma_z$).

When using this formalism, not only do we need to time-evolve the descriptors, $\mathbf{q}_a$, we also need to use time-evolved forms of transformation operators. These forms can be written in terms of their actions on the descriptors over a time-step. For example, the Hadamard gate is written in its general form as
$$ \mathbf{q}(t_1) = \left( \begin{array}{c}q_z (t)\\
-q_y(t)\\ q_x(t) \end{array} \right)$$

\noindent and the CNOT gate (with qubit 1 as control and 2 as target)
$$ \mathbf{q}_1(t_1) = \left( \begin{array}{c} q_{2x}(t)q_{1x}(t)\\q_{2x}(t)q_{1y}(t) \\  q_{1z}(t)\end{array} \right) \ \ \ 
\mathbf{q}_2 = \left( \begin{array}{c} q_{2x}(t)\\q_{2y}(t)q_{1z}(t) \\ q_{2z}(t)q_{1z}(t)\end{array} \right) $$

Two useful operators are
$$ z_{\pm} = \frac{1}{2}(\openone \pm q_z)$$

\noindent Their average values are
$$ \av{z_+} = \trace ( \rho \ketbra{0}{0} ) \ \ \ \ \ \  \ \av{z_-} = \trace ( \rho \ketbra{1}{1} )$$

\noindent That is, the average values of these operators are the diagonal elements of the density operator in the computational basis.

The use of these operators scales very easily as we add more systems. We saw above how the $q_{ai}$ for each system always describe separate subspaces for the individual system. We can therefore simply combine the $z_{\pm}$ for each system to get overall probabilities. For example, if we have two systems then
$$ p(01) = \av{z_{1+} \otimes z_{2-}}$$

\noindent which gives us one of the four diagonal elements of the 2-qubit density matrix, which are
$$(1 + \av{q_{1z}q_{z}}) \pm ( \av{q_{1z}} + \av{q_{2z}} ) , \ \ \ \ \ (1 - \av{q_{1z}q_{z}}) \pm ( \av{q_{1z}} - \av{q_{2z}} )$$

Another operator which is particularly useful to us is the density operator. It is one of the most useful ways of going between the Schr\"{o}dinger and Heisenberg pictures as in both cases we are dealing with the time-evolved operator, $\ketbra{\psi (t)}{\psi (t)}$. Therefore, unlike ordinary operators, there is no difference in the form of the density operator between the two pictures. The density operator is a proper vector in Hilbert-Schmidt space, although its evolution is different from other operators:
$$ \rho(t) = \ketbra{\psi (t)}{\psi (t)} = U \ketbra{\psi (0)}{\psi (0)} U^\dagger = U \rho(0) U^\dagger$$

The general form of a density operator in terms of the descriptors is
$$ \rho (t) = \sum_i \av{q^{(n)}_i (t)} P_{ni}  $$

\noindent where $\mathbf{q}^{(n)}$ is the $n$-dimensional set of descriptors for the system, and $\{ P_n \}$ is the $n$-dimensional Pauli group. For a single qubit system, $P_i = \sigma_i$ (where $i$ runs over four indices and $\sigma_0 = \openone$). For two qubits, $P_i = \sigma_i\otimes \sigma_j$, and the $q_i(t)$ are therefore $q_{ij} = q_{1i}\otimes q_{2j}$ (again, $q_0 = \openone$). That is,
\begin{eqnarray*} \rho_1 (t) & = & \sum_i \av{q_i (t)} \sigma_i \\
\rho_{12} (t) & = & \sum_{ij} \av{q_{1i}(t)\ q_{2j}(t)} \sigma_i \otimes \sigma_j \end{eqnarray*}

The two-qubit density operator can also be written
$$\rho (t) = \frac{1}{4} \Big( \openone \otimes \openone + \mathbf{a.\sigma} \otimes \openone + \openone \otimes \mathbf{b.\sigma} + \sum_{mn} t_{mn} \sigma_m \otimes \sigma_n \Big) $$

\noindent where
 $$ a_i = \av{q_{1i}(t)} \ ; \ \ b_i = \av{q_{2i}(t)} \ ; \ \ t_{mn} = \av{q_{1m}(t)q_{2n}(t)} $$


\section{Direct construction of Deutsch-Hayden operators}\label{const}

Using Deutsch and Hayden's method we can construct a density matrix from a set of $q_{ai}$ gained by a certain preparation procedure. For example, consider a circuit of two qubits, with a Hadamard gate operating on the first and then a CNOT with qubit 1 as control and 2 as target. The qubits start at time $t=0$ in the zero state
$$ \mathbf{q}_1 = \mathbf{\sigma} \otimes \openone \ \ \    \mathbf{q}_2 = \openone \otimes \mathbf{\sigma} $$

\noindent The Hadamard gate takes them to
$$ \mathbf{q}_1 = \left( \begin{array}{c}\sigma_z \otimes \openone \\ -\sigma_y \otimes \openone \\ \sigma_x \otimes \openone \end{array} \right) \ \ \ 
\mathbf{q}_2 = \left( \begin{array}{c}\openone \otimes \sigma_x \\ \openone \otimes \sigma_y \\ \openone \otimes \sigma_z \end{array} \right) $$
 
\noindent and then the CNOT leaves them as
\begin{equation} \mathbf{q}_1 = \left( \begin{array}{c}\sigma_z \otimes \sigma_x \\ -\sigma_y \otimes \sigma_x \\ \sigma_x \otimes \openone \end{array} \right) \ \ \ \ \ \ \ \ \ \ \
\mathbf{q}_2 = \left( \begin{array}{c}\openone \otimes \sigma_x \\ \sigma_x \otimes \sigma_y \\ \sigma_x \otimes \sigma_z \end{array} \right) \label{qiconst}\end{equation}

\noindent The density matrix is therefore
\begin{eqnarray*}\rho_{12} (t) & = & \sum_{i} \sigma_i\otimes \sigma_i\end{eqnarray*}

\noindent which is the density matrix of the state $\ket{00} + \ket{11}$.

A reasonable question to ask at this point would be: if we do not know (or do not care about) the preparation procedure, can we then do the reverse operation, and construct a set of $q_{ai}$ directly from a given density operator?

Furthermore, how unique would such a construction be? In our example of $\ket{00} + \ket{11}$, we can immediately see that the set (\ref{qiconst}) is not unique as we could swap the qubits in the preparation procedure. Performing $H_2$ then $CNOT_{2\rightarrow 1}$ would give us

\begin{equation} \mathbf{q}_1 = \left( \begin{array}{c}\sigma_x \otimes \openone \\ \sigma_y \otimes \sigma_x \\ \sigma_z \otimes \sigma_x \end{array} \right) \ \ \ \ \ \ \ \ \ \ \
\mathbf{q}_2 = \left( \begin{array}{c}\sigma_x \otimes \sigma_z \\ \sigma_x \otimes \sigma_y \\ \openone \otimes \sigma_x \end{array} \right)\label{qisym1}\end{equation}

\noindent Under what circumstances in general do different preparation procedures give rise to different sets of \q? Is this the only way that different sets of $q_{ai}$ corresponding to the same Schr\"{o}dinger state can be constructed? Essentially, how much can we trust what a given analysis of information flow tells us about the dependencies in a descriptor?
 
We start by constructing a set of $q_{ai}$ from a density matrix. The density matrix is dependent on the average values of the \q, so we must start by finding these. We can find the values of $  \av{q_{ai} (t)} $ by finding the components of $ \rho $ in the $ q_i(0) $ directions, $ \trace (\rho q_i) $. Such an operation can be simplified by noting that, writing $ \rho = \sum_n \rho_{1n} \otimes \rho_{2n} $, we have
\begin{eqnarray*} \rho_{ij} & = & \trace( \sum_n \rho_{1n} \sigma_i \otimes \rho_{2n}\sigma_j) \\
{} & = & \sum_n \trace (\rho_{1n}\sigma_i) \trace(\rho_{2n}\sigma_j)  \end{eqnarray*}

\noindent We therefore need only to look at the components in the sum where both traces are nonzero. The nonzero traces are:
\begin{eqnarray*} \trace (\ketbra{0}{0} \openone) = & \trace(\ketbra{1}{1}\openone) & = 1\\
\trace(\ketbra{0}{0} \sigma_z ) = & - \trace(\ketbra{1}{1} \sigma_z) & = 1\\
\trace(\ketbra{0}{1} \sigma_x ) = & \trace(\ketbra{1}{0} \sigma_x) & = 1\\
\trace(\ketbra{0}{1} \sigma_y ) = & - \trace(\ketbra{1}{0} \sigma_y) & = \mi \end{eqnarray*}

\noindent For example, consider the state \ket{00} + \ket{11}. The density operator is
$$ \rho = \ketbra{0}{0} \otimes \ketbra{0}{0} + \ketbra{0}{1}\otimes \ketbra{0}{1} + \ketbra{1}{0}\otimes \ketbra{1}{0} + \ketbra{1}{1}\otimes\ketbra{1}{1}$$

\noindent The nonzero components from this will be $ \openone \otimes \openone $, $ \sigma_x \otimes \sigma_x $, $ \sigma_y \otimes \sigma_y $ and $\sigma_z \otimes \sigma_z$, all of which are 1. The density matrix can therefore be written
$$ \rho = \sum_{i=0}^3 \sigma_i\otimes \sigma_i$$

\noindent which gives us
\begin{eqnarray*} \av{q_{1i}} & = & 0\\
\av{q_{2i}} & = & 0\\
\av{q_{1i}q_{2j}} & = & \delta_{ij} \end{eqnarray*}

\noindent It is instructive now to look at the simplest possible set of $q_{ai}$ that is consistent with these conditions (and also that $q_{iy} = q_{ix}q_{iz}$):
\begin{equation} \mathbf{q}_1 = \left( \begin{array}{c}\openone \otimes \sigma_x \\ \sigma_x \otimes \sigma_x \\ \sigma_x \otimes \openone \end{array} \right) \ \ \ \ \ \ \ \ \ \ \
\mathbf{q}_2 = \left( \begin{array}{c}\openone \otimes \sigma_x \\ \sigma_x \otimes \sigma_x \\ \sigma_x \otimes \openone \end{array} \right) \label{qijgiven}\end{equation}

\noindent This, however, is not a well-formed set of $q_{ai}$ as it does not form a basis in the Hilbert-Schmidt space of the two systems. A proper basis must have the following properties (\emph{cf} \cite[pp3ff]{handj}):
\begin{enumerate}
\item $N^2$ linearly independent operators: $\exists \{a_{ij}\}$ such that $\sum a_{ij} e_{ij} = 0$
\item The operators of the basis are mutually orthogonal: $\trace (e_{ij}e^\prime_{nm}) = \delta_{ij,nm}$
\item The basis operators are complete: $\trace e_{ij}^2 = 1$
\item The operators must be Hermitian
\end{enumerate}
 
\noindent 1. comes from the fact that the operators must span the entire space. 3 and 4 are consequences of the fact that all the operators are time-evolved from the $t=0$ Pauli operators, which are complete and Hermitian. This time evolution firstly cannot change the trace of an operator (3), and secondly is unitary so will take Hermitian operators to Hermitian operators.

We can see that the set (\ref{qijgiven}) fails to meet the first of these criteria: it gives only 8 linearly independent operators, rather than $4^2=16$. Contrast this with the $q_{1i}q_{2j}$ operators obtained from (\ref{qiconst}), which comprise the entire set of Dirac operators, which are known to be linearly independent and space the entire space.
 
If we now consider other $q_{ai}$ which could give a well-formed basis, we can use these criteria to note also that $\mathbf{q}_1$ and $\mathbf{q}_2$ cannot be identical: these co-ordinatise two subspaces of the whole space, and if the subspaces are isomorphic then the product of their bases will not give a basis for the whole space. Furthermore, no two $q_{ai}$ can be identical, and neither can any two $q_{1i}q_{2j}$. In the first case, the resultant product operator would be $\openone \otimes \openone$, which is already given by $q_{10}q_{20}$ ($q_0 = \openone$ is always understood), and hence in both instances there would be fewer than 16 linearly independent operators. 

Let us see now how (\ref{qijgiven}) may be altered to make it well-formed. The easiest way of making the combined operators span the system is to introduce $\sigma_y$ instead of one $\sigma_x$ (preserving the average values) and $\sigma_z$ for a $\openone$:
$$ \mathbf{q}_1 = \left( \begin{array}{c}\openone , \sigma_z
\otimes \sigma_x , \sigma_y \\  q_{1x}q_{1z}\\ \sigma_x , \sigma_y \otimes \openone , \sigma_z \end{array} \right) \ \ \ \ \ \ \ \ \ \ \
\mathbf{q}_2 = \left( \begin{array}{c}\sigma_z , \openone \otimes \sigma_y , \sigma_x \\ q_{2x}q_{2z} \\ \sigma_x , \sigma_y \otimes \openone \sigma_z  \end{array} \right) $$

\noindent Only two of these sets of these operators span the entire space:
$$ \mathbf{q}_1 = \left( \begin{array}{c}\sigma_z \otimes \sigma_x \\ -\sigma_y \otimes \sigma_x \\ \sigma_x \otimes \openone \end{array} \right) \ \ \ \ \ \ \ \ \ \ \
\mathbf{q}_2 = \left( \begin{array}{c}\openone \otimes \sigma_x \\ \sigma_x \otimes \sigma_y \\ \sigma_x \otimes \sigma_z \end{array} \right) $$

\noindent and
\begin{equation} \mathbf{q}_1 = \left( \begin{array}{c}\openone \otimes \sigma_x \\ \sigma_x \otimes \sigma_y \\ \sigma_x \otimes \sigma_z \end{array} \right) \ \ \ \ \ \ \ \ \ \ \
\mathbf{q}_2 = \left( \begin{array}{c}\sigma_z \otimes \sigma_x \\ -\sigma_y \otimes \sigma_x \\ \sigma_x \otimes \openone \end{array} \right)\label{swapHH}\end{equation}

We now have three different sets of $q_{ai}$ corresponding to the same Schr\"{o}dinger state: (\ref{qiconst}), (\ref{qisym1}) and (\ref{swapHH}). The interesting question now is to what extent such sets are unique -- in the present case we have found three possible sets, are there any more that would represent the same state? In general, how many distinct sets would we be able to find?

In order to answer this question we will first look at how many unknowns there are in the system (that is, how many unknown quantities are needed fully to determine a set of \q), and then consider how many constraints there are on their values. We start by writing a general form of a $q_i$ on the $a$th system:
\begin{equation} q_{ai} = \sum_{k,l=0}^3 a^{(i)}_{kl} \sigma_k\otimes \sigma_l \label{qghn}\end{equation}

\noindent We know that $q_{10} = q_{20} = \openone$, so the general forms for $q_{1i}\otimes q_{2j}$ are ($n,m \in \{x,y,z\}$)
\begin{eqnarray*} q_{10}q_{20} & = & \openone \otimes \openone \\
q_{1i}q_{20} & = & \sum_{k,l=0}^3 a^{(i)}_{kl} \sigma_k \otimes \sigma_l\\
q_{10}q_{2j} & = & \sum_{m,n}^3 b^{(j)}_{mn} \sigma_m \otimes \sigma_n\\
q_{1i}q_{2j} & = & \sum_{k,l,m,n} a^(i)_{kl}b^{(j)}_{mn} \sigma_k \sigma_m \otimes \sigma_l \sigma_n\end{eqnarray*}

\noindent We can see from this that if we can find the $a^{(i)}_{kl}$ and $b^{(j)}_{mn}$ then we will have fully described the set of operators. $a_{kl}$ for a given $i$ and $b_{mn}$ for a given $j$ each contain 16 unknowns ($k,l,m,n \in \{0,x,y,z\}$). Now we know that $q_{ay} = q_{ax}q_{az}$, so for each system we have $2(16) = 32$ unknowns. Therefore, for our system here we will need 64 numbers fully to determine a set of \q.

So how many constraints do we have for the system? First, we have the conditions that no two $q_{ai}$ are the same, and that they each have norm 1:
\begin{eqnarray*} \trace (q_{1i}q_{1j} ) & = & \delta_{ij}\\
\trace (q_{2i}q_{2j} ) & = & \delta_{ij}\\
\trace (q_{1i}q_{2j}) & = & 0 \end{eqnarray*}

\noindent where $i,j \in \{0,x,z\}$. Now, $q_{a0}$ is fixed at $\openone \otimes \openone$, so in each of the above equations the case $n=m=0$ tells us nothing new. Therefore we have $3(9-1) = 24$ constraint equations here.

Next we have the constraint that each $q_{ai}$, $i\neq 0$, must be traceless. To see why, consider two Hilbert-Schmidt spaces, one co-ordinatised by the Dirac basis $\{\sigma_i \otimes \sigma_j\}$, and the other by $\{q_{1i}\otimes q_{2j}\}$. There exists a linear transformation between the two, given by $T$. Because both spaces are of Hermitian operators, we require $T$ to be unitary to preserve hermicity. Therefore, the elements of $\{q_{1i}\otimes q_{2j}\}$ are a unitary transformation of the elements of $\{\sigma_i \otimes \sigma_j\}$. The trace operation is invariant under such transformation, so the elements $q_{1i}\otimes q_{2j}$ must have the same values for their trace as $\sigma_i \otimes \sigma_j$ -- that is, $\delta_{i0}\delta_{j0}$. This gives us the four constraint equations
$$ a^{(n)}_{00} =b^{(m)}_{00} = 0$$

Another set of constraints comes from the criterion that the elements of each $\mathbf{q}_a$ are linearly independent. That is, that there exist some constants $c_n$ for which
$$ \sum_i c_i q_{1i} = 0$$

\noindent (and for $q_{2j}$). Substituting (\ref{qghn}) into this equation we have
$$\sum_{ikl} c_i a^{(i)}_{kl} \sigma_k \otimes \sigma_l = 0$$

\noindent We know that the Dirac operators are mutually linearly independent, so each $(k,l)$ term in the above sum must equal zero independently:
$$ \sum_i c_i a^{(i)}_{kl} = 0 \ \ \ \ \ \forall k,l$$

\noindent As $k,l \in \{0,x,y,z\}$ this gives us 16 equations for each system. However, we already know that $a^{(i)}_{00} = b^{(m)}_{00} = 0$, so we have a total of $2(16) - 4 = 28$ constraints here.

It is interesting to note that up to this point, the constraints on the values of $q_{ai}$ have come exclusively from the structure of the Hilbert-Schmidt space - these are constraints on \emph{any} physical system. We now move on to constrain our operators to a particular physical state. These constraints are the elements of the density matrix: 
\begin{equation} \rho_{ij} = \av{q_{1i}q_{2j}} = \trace ( q_{1i}q_{2j} \ketbra{0}{0} )\label{symrhp}\end{equation}

\noindent This gives us nine equations, but only eight constraints as we know $\av{q_{10}q_{20}}$ already (1). 

This gives us a total of $24+4+28+8 = 64$ constraint equations for 64 unknown variables. Thus we see that, in general, the structure of the $q_{ai}$ is fully determined by the physical system. That is, \textbf{in general a given state defined by a density matrix has a unique representation in terms of Deutsch-Hayden operators}. 

The caveat ``in general" evidently applies as we have been dealing with a situation where the representation is \emph{not} unique. Why is this the case for our above example $\ket{00} + \ket{11}$? The reason is that we have a physical symmetry between the two systems, and this is picked up in the mathematics. All of the `structural' constraints are, of necessity, symmetric between the systems and between basis states within each system (or else they would be making a physical statement). The physical symmetries are all contained within the structure of the density matrix. We can see from (\ref{symrhp}) that any symmetries of the density matrix will give rise to different sets of $q_{ai}$ corresponding to those symmetries. That is, if there is a symmetry of the density operator pertaining to the element $\rho_{ij}$, then the same symmetry applied to $q_{1i}q_{2j}$ will give us a physical state of affairs indistinguishable from the original. (\ref{symrhp}) tells us that such symmetries are the \emph{only} means of generating separate sets of  $q_{ai}$s.

Let us look at the density matrix in terms of the $\{ \sigma_i\otimes \sigma_j\}$:
$$ \rho = \sum_{i,j=0}^3 a_{ij} \sigma_i\otimes \sigma_j$$

\noindent The symmetries that will give us different sets of $q_{ai}$ are the symmetries of $a_{ij}$:
$$ \left( \begin{array}{cccc} 1 & 0 & 0 & 0 \\ 0 & 1 & 0 & 0 \\ 0 & 0 & 1 & 0\\ 0 & 0 & 0 & 1\end{array} \right)$$


It is important to note at this point that we cannot have any symmetry that violates the constraints that we have imposed on the system above: so, for example, we cannot swap $q_{1x}$ with another $q_{ai}$ whose product with $q_{1z}$ is nonzero. Furthermore, $q_{1i}$ cannot be swapped with $q_{1i}q_{2j}$ unless $j=0$: otherwise it would force $q_{2j} = \openone \otimes \openone$ which would give less than 16 independent operators. To put that all in other words, we are restricted to the \emph{physical} symmetry transformations of the system, non-physical transformations violating the constraints we have already laid on the system. In terms of symmetries of the density matrix, we are therefore restricted to swapping rows and columns.

\noindent The symmetries of the density matrix are as follows ($\left\{ \right\} $ denotes a single transformation). Firstly, swapping single rows and the same numbered columns (\emph{ie} reflections in the diagonal):
\begin{eqnarray*} \left\{ \openone \otimes \openone  \leftrightarrow \openone \otimes \openone \right\} & \ \ \ \ & \left\{ \sigma_x \otimes \openone  \leftrightarrow \openone \otimes \sigma_x \right\} \\ 
\left\{ \sigma_y \otimes \openone  \leftrightarrow \openone \otimes \sigma_y \right\} & \ \ \ \ & \left\{ \sigma_z \otimes \openone  \leftrightarrow \openone \otimes \sigma_z \right\}\end{eqnarray*}

\noindent Then swapping two rows and two columns at a time:
\begin{eqnarray*} r_1 \leftrightarrow r_2 \ \mathrm{and} \ c_1 \leftrightarrow c_2 & : & \ \  \left\{ \begin{array}{c} \openone \otimes \openone \leftrightarrow \openone \otimes \sigma_x \\ \openone \otimes \openone \leftrightarrow \sigma_x \otimes \openone \end{array}\right\} \\
r_1 \leftrightarrow r_3 \ \mathrm{and} \ c_1 \leftrightarrow c_3 & : & \ \  \left\{ \begin{array}{c} \openone \otimes \openone \leftrightarrow \openone \otimes \sigma_y \\ \openone \otimes \openone \leftrightarrow \sigma_y \otimes \openone \end{array}\right\} \\
r_1 \leftrightarrow r_4 \ \mathrm{and} \ c_1 \leftrightarrow c_4 & : & \ \  \left\{ \begin{array}{c} \openone \otimes \openone \leftrightarrow \openone \otimes \sigma_z \\ \openone \otimes \openone \leftrightarrow \sigma_z \otimes \openone \end{array}\right\} \\
r_2 \leftrightarrow r_3 \ \mathrm{and} \ c_2 \leftrightarrow c_3 & : & \ \  \left\{ \begin{array}{c} \openone \otimes \sigma_x \leftrightarrow \openone \otimes \sigma_y \\ \sigma_x \otimes \openone \leftrightarrow \sigma_y \otimes \openone \end{array}\right\} \\
r_2 \leftrightarrow r_4 \ \mathrm{and} \ c_2 \leftrightarrow c_4 & : & \ \  \left\{ \begin{array}{c} \openone \otimes \sigma_x \leftrightarrow \openone \otimes \sigma_z \\ \sigma_x \otimes \openone \leftrightarrow \sigma_z \otimes \openone \end{array}\right\} \\
r_3 \leftrightarrow r_4 \ \mathrm{and} \ c_2 \leftrightarrow c_4 & : & \ \  \left\{ \begin{array}{c} \openone \otimes \sigma_y \leftrightarrow \openone \otimes \sigma_z \\ \sigma_y \otimes \openone \leftrightarrow \sigma_z \otimes \openone \end{array}\right\} \end{eqnarray*}

\noindent Any combinations of the above will also be symmetries of the matrix.

We can immediately discard symmetries of the form $\openone \otimes \openone \leftrightarrow \openone \otimes \sigma_z$: these will merely change the position of $\openone \otimes \openone$ in a set of \q, but we always have it as $q_{a0}$ by convention. We can also discard several as $q_y$ is not independent of $q_x$ and $q_z$ - for example, swapping $r_2 \leftrightarrow r_3 \ \mathrm{and} \ c_2 \leftrightarrow c_3$ and swapping $r_2 \leftrightarrow r_4 \ \mathrm{and} \ c_2 \leftrightarrow c_4$ imply the same thing. We are therefore left with the following symmetries on a set of \q:
\begin{equation} q_{1x} \leftrightarrow q_{2x} \label{sim1}  \end{equation}
\begin{equation} q_{1y} \leftrightarrow q_{2y}\label{sim4} \end{equation}
\begin{equation} q_{1z} \leftrightarrow q_{2z}\label{sim2} \end{equation}
\begin{equation} \left\{ \begin{array}{c} q_{1x} \leftrightarrow q_{1z} \\ q_{2x} \leftrightarrow q_{2z} \end{array}\right\} \label{sim3} \end{equation}
\begin{equation} \left\{ \begin{array}{c} q_{1x} \leftrightarrow q_{1y} \\ q_{2x} \leftrightarrow q_{2y} \end{array}\right\} \label{sim5} \end{equation}
\begin{equation} \left\{ \begin{array}{c} q_{1y} \leftrightarrow q_{1z} \\ q_{2y} \leftrightarrow q_{2z} \end{array}\right\} \label{sim6} \end{equation}

\noindent We also have the combinations of the above (neglecting duplicates, and also leaving out the third in any given set of $q_i$ as it is implied by the first two):
\begin{equation} \left\{ \begin{array}{c} q_{1x} \leftrightarrow q_{2x}\\ q_{1z} \leftrightarrow q_{2z} \end{array}\right\} \label{sim7} \end{equation}
\begin{equation} \left\{ \begin{array}{c} q_{2x} \leftrightarrow q_{1z} \\ q_{1x} \leftrightarrow q_{2z}\end{array}  \right\} \label{sim10} \end{equation}
\begin{equation} \left\{ \begin{array}{c} q_{2x} \leftrightarrow q_{1y} \\ q_{1x} \leftrightarrow q_{2y}\end{array} \right\} \label{sim8} \end{equation}
\begin{equation} \left\{ \begin{array}{c} q_{2z} \leftrightarrow q_{1y} \\ q_{1z} \leftrightarrow q_{2y}\end{array} \right\} \label{sim9} \end{equation}

\noindent These are all the symmetries present: if we take further combinations of (\ref{sim7}-\ref{sim9}) with (\ref{sim1}-\ref{sim6}) we get a symmetry already given. Therefore, the identity transformation $\q \leftrightarrow \q$ plus the transformations (\ref{sim1}) - (\ref{sim9}) form the group of relevant symmetries of the density matrix as applied to the $q_{ai}$ operators.
 
We can now use this group to generate the set of sets of $q_{ai}$ that will correspond to the state \ket{00} + \ket{11}. Now, because the only way to generate a different set of $q_{ai}$ is by one of these symmetry transformations, each element in the set of sets of $q_{ai}$ will have orbit 1 when acted on by the group of symmetries (we denote this by $G_{\rho}$) -- that is, any element can be reached by any other using one of the group elements. For example, (\ref{qisym1}) is generated from (\ref{qiconst}) by applying (\ref{sim3}), and (\ref{swapHH}) by applying (\ref{sim4}). We can therefore use (\ref{qiconst}) to generate the entire set of allowable sets of \q:
$$ \mathbf{q}^{(1)}_1 = \left( \begin{array}{c}\sigma_z \otimes \sigma_x \\ -\sigma_y \otimes \sigma_x \\ \sigma_x \otimes \openone \end{array} \right) \ \ \ \ \ \ \ \ \ \ \
\mathbf{q}^{(1)}_2 = \left( \begin{array}{c}\openone \otimes \sigma_x \\ \sigma_x \otimes \sigma_y \\ \sigma_x \otimes \sigma_z \end{array} \right) $$
$$ \mathbf{q}^{(2)}_1 = \left( \begin{array}{c}\openone \otimes \sigma_x \\ \sigma_x \otimes \sigma_x \\ \sigma_x \otimes \openone \end{array} \right) \ \ \ \ \ \ \ \ \ \ \
\mathbf{q}^{(2)}_2 = \left( \begin{array}{c}\sigma_z \otimes \sigma_x \\ -\sigma_y \otimes \sigma_y \\ \sigma_x \otimes \sigma_z \end{array} \right) $$
$$ \mathbf{q}^{(3)}_1 = \left( \begin{array}{c}\openone \otimes \sigma_y \\ \sigma_x \otimes \sigma_y \\ \sigma_x \otimes \openone \end{array} \right) \ \ \ \ \ \ \ \ \ \ \
\mathbf{q}^{(3)}_2 = \left( \begin{array}{c}\sigma_z \otimes \sigma_y \\ -\sigma_y \otimes \sigma_x \\ \sigma_x \otimes \sigma_z \end{array} \right) $$
$$ \mathbf{q}^{(4)}_1 = \left( \begin{array}{c}\sigma_z \otimes \sigma_x \\ \sigma_x \otimes \sigma_y \\ -\sigma_y \otimes \sigma_z \end{array} \right) \ \ \ \ \ \ \ \ \ \ \
\mathbf{q}^{(4)}_2 = \left( \begin{array}{c}\openone \otimes \sigma_x \\ -\sigma_y \otimes \sigma_x \\ -\sigma_y \otimes \openone \end{array} \right) $$
$$ \mathbf{q}^{(5)}_1 = \left( \begin{array}{c}\sigma_z \otimes \sigma_x \\ -\sigma_y \otimes \sigma_y \\ \sigma_x \otimes \sigma_z \end{array} \right) \ \ \ \ \ \ \ \ \ \ \
\mathbf{q}^{(5)}_2 = \left( \begin{array}{c}\openone \otimes \sigma_x \\ \sigma_x \otimes \sigma_x \\ \sigma_x \otimes \openone \end{array} \right) $$
$$ \mathbf{q}^{(6)}_1 = \left( \begin{array}{c}\sigma_x \otimes \openone \\ \sigma_y \otimes \sigma_x \\ \sigma_z \otimes \sigma_x \end{array} \right) \ \ \ \ \ \ \ \ \ \ \
\mathbf{q}^{(6)}_2 = \left( \begin{array}{c}\sigma_x \otimes \sigma_z \\ -\sigma_x \otimes \sigma_y \\ \openone \otimes \sigma_x \end{array} \right) $$
$$ \mathbf{q}^{(7)}_1 = \left( \begin{array}{c}-\sigma_y \otimes \sigma_x \\ \sigma_z \otimes \sigma_x \\ \sigma_x \otimes \openone \end{array} \right) \ \ \ \ \ \ \ \ \ \ \
\mathbf{q}^{(7)}_2 = \left( \begin{array}{c}\sigma_x \otimes \sigma_y \\ \openone \otimes \sigma_x \\ \sigma_x \otimes \sigma_z \end{array} \right) $$
$$ \mathbf{q}^{(8)}_1 = \left( \begin{array}{c}\sigma_z \otimes \sigma_x \\ \sigma_x \otimes \openone \\ -\sigma_y \otimes \sigma_x \end{array} \right) \ \ \ \ \ \ \ \ \ \ \
\mathbf{q}^{(8)}_2 = \left( \begin{array}{c}\openone \otimes \sigma_x \\ \sigma_x \otimes \sigma_z \\ \sigma_x \otimes \sigma_y \end{array} \right) $$
$$ \mathbf{q}^{(9)}_1 = \left( \begin{array}{c}\openone \otimes \sigma_x \\ \sigma_x \otimes \sigma_y \\ \sigma_x \otimes \sigma_z \end{array} \right) \ \ \ \ \ \ \ \ \ \ \ 
\mathbf{q}^{(9)}_2 = \left( \begin{array}{c}\sigma_z \otimes \sigma_x \\ -\sigma_y \otimes \sigma_x \\ \sigma_x \otimes \openone \end{array} \right) $$
$$ \mathbf{q}^{(10)}_1 = \left( \begin{array}{c}\sigma_x \otimes \sigma_z \\ -\sigma_x \otimes \sigma_y \\ \openone \otimes \sigma_x \end{array} \right) \ \ \ \ \ \ \ \ \ \ \
\mathbf{q}^{(10)}_2 = \left( \begin{array}{c}\sigma_x \otimes \openone \\ \sigma_y \otimes \sigma_x \\ \sigma_z \otimes \sigma_x \end{array} \right) $$
$$ \mathbf{q}^{(11)}_1 = \left( \begin{array}{c}\sigma_x \otimes \sigma_y \\ \openone \otimes \sigma_x \\ \sigma_x \otimes \sigma_z \end{array} \right) \ \ \ \ \ \ \ \ \ \ \
\mathbf{q}^{(11)}_2 = \left( \begin{array}{c}-\sigma_y \otimes \sigma_x \\ \sigma_z \otimes \sigma_x \\ \sigma_x \otimes \openone \end{array} \right) $$
$$ \mathbf{q}^{(12)}_1 = \left( \begin{array}{c}\openone \otimes \sigma_x \\ \sigma_x \otimes \sigma_z \\ \sigma_x \otimes \sigma_y \end{array} \right) \ \ \ \ \ \ \ \ \ \ \
\mathbf{q}^{(12)}_2 = \left( \begin{array}{c}\sigma_z \otimes \sigma_x \\ \sigma_x \otimes \openone \\ -\sigma_y \otimes \sigma_x \end{array} \right) $$

There is one final question that needs to be answered about these alternative sets of \q: do they give the correct results when they subject to further evolution of the system? Do all these sets give the same average values after arbitrary unitary evolution?

It is fairly straightforward to prove that this is the case. Consider two set of operators sharing the same average values, $q_{ai}$ and $q^\prime_{ai}$. $q_{ai}$ has been constructed from a circuit and so is known to give the correct evolution, \emph{ie} $\rho(t_1) = \sum_{ij} \av{q_{1i}(t)q_{2j}(t)} \sigma_i \otimes \sigma_j$. Because $q_{ai}$ and $q^\prime_{ai}$ share the same average values, we can write
$$ \rho(t) = \sum_{ij} \av{q^\prime_{1i}q^\prime_{2j}} \sigma_i \otimes \sigma_j$$

\noindent Let $U$ be the transformation taking place between $t$ and $t_1$:
\begin{eqnarray*} \rho (t_1) & = & U \rho(t) U^\dagger \\
 {} & = & \sum_{ij} \av{q^\prime_{1i}q^\prime_{2j}} \ U \sigma_i \otimes \sigma_j U^\dagger \\
 {} & = & \sum_{ij} \av{q^\prime_{1i}q^\prime_{2j}} \sum_{ab} \trace ( \sigma_a \otimes \sigma_b \   U \sigma_i \otimes \sigma_j U^\dagger ) \ \sigma_a \otimes \sigma_b \end{eqnarray*}
 
 \noindent We showed above how any basis can be written as $T^\dagger \sigma_n \otimes \sigma_m T$:
\begin{eqnarray*} \rho (t_1)  & = &  \sum_{ij} \av{ U^\dagger (t-t_0) \sigma_i \otimes \sigma_j U(t-t_0)} \\
& & {}. \sum_{ab} \trace ( \sigma_a \otimes \sigma_b \   U \sigma_i \otimes \sigma_j U^\dagger ) \ \sigma_a \otimes \sigma_b \end{eqnarray*}
 
\noindent Now
\begin{eqnarray*} \trace (\sigma_a \otimes \sigma_b \   U \sigma_i \otimes \sigma_j U^\dagger ) & = & \trace ( UU^\dagger \sigma_a \otimes \sigma_b \   U \sigma_i \otimes \sigma_j U^\dagger )\\
{} & = & \trace (U^\dagger \sigma_a \otimes \sigma_b \   U \sigma_i \otimes \sigma_j )\\
{} & = & \left \{ \begin{array}{l} 1 \ \ \ \ \ \mathrm{if} \ \sigma_i \otimes \sigma_j = U^\dagger \sigma_a \otimes \sigma_b U \\ 0 \ \ \ \ \ \mathrm{otherwise} \end{array} \right. \end{eqnarray*}

\noindent Therefore
 \begin{eqnarray*} \rho (t_1) & = &  \sum_{ab} \av{ U^\dagger (t-t_0) U^\dagger\sigma_a \otimes \sigma_b U U(t-t_0)} 
\ \sigma_a \otimes \sigma_b \\
{} & = & \sum_{ab} \av{ U^\dagger (t_1-t_0) \sigma_a \otimes \sigma_b U(t_1-t_0)} \ \sigma_a \otimes \sigma_b\\
{} & = & \sum_{ab}\av{ U^\dagger (t_1-t) q^\prime_{1a}(t)q^\prime_{2b}(t) U(t_1 - t) } \ \sigma_a \otimes \sigma_b\\
{} & = & \sum_{ab}\av{q^\prime_{1a}(t_1)q^\prime_{2b}(t_1)} \sigma_a \otimes \sigma_b \end{eqnarray*}

\noindent In other words, for arbitrary evolution, the evolved density operator can be written in terms of the evolved $q^\prime$ operators, and 
$$\av{q^\prime_{1i}(t_1)q^\prime_{2j}(t_1)} = \av{q_{1i}(t_1)q_{2j}(t_1)}$$
 
We have therefore shown that in the Deutsch-Hayden picture a given Schr\"{o}dinger state is represented by the $G_{\rho}$-set of sets of $q_{ai}$ which is acted on by the group $G_{\rho}$ of symmetries of the density matrix $\rho$ of the state. Each element of the $G_{\rho}$-set has orbit 1, and for all practical purposes any element may be used in place of any other. We therefore see that in an analysis of information flow, the possible descriptors for the systems under consideration differ only by the physical symmetries of the systems -- that is, the description of information flow and dependencies will not differ in physically significant ways between different sets of descriptors for the same situation. Any set that we chose will give us the correct description.

\section{The measurement interaction}\label{minter}

We now turn to measurement-type interactions in terms of the Deutsch-Hayden picture. The simplest form of a measurement interaction is full decoherence, where the system is measured by an ancillary system of the same size to which we do not have access. The simplest model of this interaction is that there is a CNOT gate with the system as the control and the ancilla as target. For example, consider a one-qubit system and a one-qubit ancilla. Without loss of generality, we can take the state of the ancilla before interaction to be $\ket{\mathbf{0}} \Rightarrow \mathbf{q}_a = \openone \otimes \mathbf{\sigma}$. The state of the system before is given by $\mathbf{q}_s(t)$. The state of the system afterwards corresponds to
$$ \mathbf{q}_s = \left( \begin{array}{c}q_{sx}(t) \otimes \sigma_x \\ q_{sy}(t) \otimes \sigma_x \\ q_{sz}(t) \otimes \openone \end{array} \right) $$

\noindent That is, the only nonzero $\av{q_{si}}$ is $q_{sz}$. Now, the diagonal elements of the density matrix are $\openone \pm \sigma_z$ -- so what we have is the standard action of decoherence, where the off-diagonal elements of the density matrix written in the decoherence basis go to zero and the diagonal elements are unchanged. 

We can extend this to any size of system if we have an ancilla at least as large as the system being decohered. Then each qubit of the system will perform a CNOT with a given qubit of the ancilla, leaving only the $\av{q_{nz}}$ as nonzero. In any size system, the diagonal elements are given by the coefficients of combinations of $\openone$ and $\sigma_z$'s -- that is, combinations of $\av{\openone}$ and $\av{q_z}$ only. Thus again we have the off-diagonal elements becoming zero and the diagonal elements unchanged.

Moving away from decoherence, let us now suppose that we have access to the ancilla. In our two-system example, the operators for this ancilla will be
$$ \mathbf{q}_a = \left( \begin{array}{c}\openone \otimes \sigma_x \\ q_{sz}(t) \otimes \sigma_y \\ q_{sz}(t) \otimes \sigma_z \end{array} \right) $$

\noindent We can see from this that only the information contained within the $q_{sz}(t)$ operator is picked up by the ancilla. This is because, were we at a later time to have access only to the ancilla, the information about the first system that could be extracted is $q_{sz}(t)$ only. Again, this is what we want: the $q_{sz}(t)$ components give the probabilities of outcomes of measurement in the measurement basis (as they give the diagonal elements of the density matrix). This is all the information about a state that can be gained by measuring it.

We have seen what happens when we perform a measurement in the \ket{0},\ket{1}  basis. What if we wish to measure in a different basis? In such a case, the system would still only have nonzero $\av{q_z}$ component (and the ancilla only pick up information from $q_z$) -- but \emph{after} the $q_{ai}$ had been rotated in the change of basis. For example, suppose we wished to measure the system in the basis $\ket{0}\pm\ket{1}$. To change to this basis from \ket{0}, \ket{1} we perform a Hadamard transformation. The system operators are therefore $(\sigma_z, -\sigma_y, \sigma_x)$. Performing a CNOT with the ancilla then gives us
$$ \mathbf{q}_s = \left( \begin{array}{c}q_{sz}(t) \otimes \sigma_x \\ -q_{sy}(t) \otimes \sigma_x \\ q_{sx}(t) \otimes \openone \end{array} \right) $$

\noindent That is, the coefficient of the non-zero component $\sigma_z$ is $\av{q_{sx}(t)}$.

What we have in effect done here is change the meaning of $\mathbf{\sigma}$ as the basis changes, and compensate by changing the coefficients. $\mathbf{\sigma}$ becomes with respect to the new basis rather than \ket{0}, \ket{1} and the coefficients will necessarily change.

In modelling the measurement of the system as a CNOT gate with an ancilla as the target, we have been dealing solely with projective measurements. For completeness, we would like to look at the case of generalised measurement, where the states of the measurement basis are not necessarily mutually orthogonal.

A POVM measurement can always be modelled by adding a second system to the one being measured, performing a projective measurement on both systems and then tracing out the second system. Now, we have seen that the density operator \emph{in any basis} can always be written as
\begin{equation} \openone \otimes \openone + \mathbf{a.\sigma} \otimes \openone + \openone \otimes \mathbf{b.\sigma} + \sum_{mn} t_{mn} \sigma_m \otimes \sigma_n \label{laksoi}\end{equation}

\noindent When the second system is traced out this becomes
$$ \openone + \mathbf{a.\sigma}$$

\noindent All the interesting information is therefore contained in $\mathbf{a} = q_{1z}$.

Take the single-qubit system we measured above. We add a second system, choosing it to be in the \ket{0} state. Now we measure the joint system in the Bell basis. In terms of the $q_{ai}$ this means performing a Hadamard then a CNOT gate. At the end of this procedure the operators for the two systems are:
$$ \mathbf{q}_1 = \left( \begin{array}{c}q_{1z}(t) \otimes \sigma_z \\ -q_{1y}(t) \otimes \sigma_y \\ q_{1x} \otimes \openone \end{array} \right) \ \ \ \ \ \ \ \ \ \ \ \mathbf{q}_2 = \left( \begin{array}{c}\openone \otimes \sigma_x \\ q_{1y}(t) \otimes \sigma_y \\ q_{1x}(t) \otimes \sigma_z \end{array} \right)$$

\noindent To complete the measurement, we must add a two-qubit ancilla and perform CNOTs between the two systems and the ancilla. We saw above that this will make all the averages for the two qubits individually zero except $\av{q_z}$. Therefore, in the generalised as in the projective measurement, what can be gained at measurement is knowledge of $q_{sz}$.

In what way, then, is the generalised measurement more powerful than a projective measurement? In terms of the Deutsch-Hayden operators, the difference comes in the choice of the basis for measurement. There are a much greater range of operations that can be carried out on two systems rather than one, and so there is more opportunity to as it were `move' different information into the $q_{z}$ `slot' from where it can be picked up by the measurement.

\section{Relative states}\label{relssstates}

We have seen what happens to the state of a system which is measured, and where the ancilla which performed the measurement is subsequently traced out. We now look at what happens when the ancilla is not ignored, but rather one state of it is singled out -- in such a case, what is the relative state of the system that was measured, as given by the Deutsch-Hayden descriptors? The above results will, of course, be the special case where the relative state of the ancilla is completely unknown -- that is, $\openone$. 

As is well known, if we have two systems with joint density matrix $\rho_{12}$ and a state of the second system $\ketbra{\beta}{\beta}$ (this can be one of the elements of a POVM), then the state of the first system is given by the partial trace
$$ \rho_1 = \trace_1 ( \rho_{12} \ketbra{\beta}{\beta} )$$

\noindent Writing the density matrix as (\ref{laksoi}) we have
\begin{eqnarray*} \rho_1 & = & \openone . \bra{\beta} \openone \ket{\beta} + \mathbf{a.\sigma}. \bra{\beta} \openone \ket{\beta} + \openone. \bra{\beta} \mathbf{b.\sigma} \ket{\beta}  \\
& & {} + \sum_{mn} t_{mn} \sigma_m. \bra{\beta} \sigma_n \ket{\beta}  \\
{} & = & \openone ( 1 + \bra{\beta} \mathbf{b.\sigma} \ket{\beta} ) + \mathbf{a.\sigma}  \\
& & {} + \sum_{mn} t_{mn}\bra{\beta} \sigma_n \ket{\beta} \sigma_m \end{eqnarray*}

\noindent We want to write this in the form $\sum_i a_i^\prime \sigma_i$, which gives us
$$ a_i^\prime = a_i + \sum_n t_{in} \bra{\beta} \sigma_n \ket{\beta}$$

\noindent In terms of the $q_{ai}$ we therefore have
$$ \av{q_{1i}} \longrightarrow \av{q_{1i}} + \av{ \sum_n q_{1i}q_{2n} \bra{\beta} \sigma_n \ket{\beta}}$$

\noindent This can be achieved if the $q^\prime_i$ for the first system relative to the state $\ketbra{\beta}{\beta}$ of the second system are
$$ q^\prime_{1i} = q_{1i} \Big( 1 + \sum_n \av{\sigma_n}_{\ketbra{\beta}{\beta}} q_{2n} \Big)$$ 

\noindent This expression is useful if we have the relative Schr\"{o}dinger state. We can also frame this is terms of relative Deutsch-Hayden operators. If $q^\prime_{2i}$ are the operators for the second system when it is in the state \ket{\beta} then we know that
$$ \av{\sigma_n}_{\ketbra{\beta}{\beta}} = \av{q^\prime_{2n}}$$

\noindent (\emph{ie} the coefficient of $\sigma_n$ in the expansion of the density operator). The operators for system 1, $q^\prime_{1i}$, relative to the operators $q^\prime_{2i}$ of the second system, are (unprimed operators are for the original state of the systems)
\begin{equation} q^\prime_{1i} = q_{1i} \Big( 1 + \sum_n \av{q^\prime_{2n}} \ q_{2n} \Big)\label{jfda}\end{equation}

\noindent We check this for the case where the second system is simply discarded, and its density matrix is therefore $\openone$. In that case we have (remembering that $n$ runs from 1 to 3)
\begin{eqnarray*} q^\prime_{1i} & = & q_{1i} \Big( 1 + \sum_n \av{q^\prime_{2n}} \ q_{2n} \Big)\\
{} & = & q_{1i} ( 1 + 0)\\
{} & = & q_{1i} \end{eqnarray*}

\noindent which is as we found above.

We can now use the expression for relative states to look in further detail at the measurement interaction. Let us consider here a specific example, where the system to be measured is in the state $\ket{0} + \ket{1}$, with descriptors
$$  \mathbf{q}_1 = \left( \begin{array}{c}\sigma_z \otimes \openone \\ -\sigma_y \otimes \openone \\ \sigma_x \otimes \openone \end{array} \right)$$

\noindent Now we add the ancilla and perform the CNOT operation:
\begin{equation}  \mathbf{q}_1 = \left( \begin{array}{c}\sigma_z \otimes \sigma_x \\ -\sigma_y \otimes \sigma_x \\ \sigma_x \otimes \openone \end{array} \right) \ \ \ \ \ \ \ \ \ \ \ \mathbf{q}_2 = \left( \begin{array}{c}\openone \otimes \sigma_x \\ \sigma_x \otimes \sigma_y \\ \sigma_x \otimes \sigma_z \end{array} \right) \label{psapihy}\end{equation}

\noindent Now we can look at the states of the first system relative to those of the ancilla. We choose the relative states of the ancilla to be in the computational basis, and use (\ref{jfda}) to give us the relative state of the system. The relative operators of the system are therefore
\begin{equation}  \mathbf{q}^\prime_1 = \left( \begin{array}{c}\sigma_z \otimes \sigma_x - \sigma_y\otimes \sigma_y \\ -\sigma_y \otimes \sigma_x + \sigma_z \otimes \sigma_y \\ \sigma_x \otimes \openone + \openone \otimes \sigma_z \end{array} \right)\label{prone}\end{equation}
\begin{equation}  \mathbf{q}^{\prime\prime}_1 = \left( \begin{array}{c}\sigma_z \otimes \sigma_x + \sigma_y\otimes \sigma_y \\ -\sigma_y \otimes \sigma_x - \sigma_z \otimes \sigma_y \\ \sigma_x \otimes \openone - \openone \otimes \sigma_z \end{array} \right)\label{prtwo}\end{equation}

\noindent where the primed are relative to \ketbra{0}{0} and the double-primed to \ketbra{1}{1}. We can see from these $q_{ai}$ that the states of the systems in both cases are eigenstates of one of the $z_\pm$ operators (recall that these operators give the probabilities for 0 and 1 in the computational basis). An interesting question at this point is to ask what the corresponding operators for the ancilla are -- that is, what are the complete sets of primed and double-primed operators? 

To answer this question we must pay close attention to the physical situation of the measurement. We have in a rather cavalier fashion introduced the states \ketbra{0}{0} and \ketbra{1}{1} of the ancilla, which is part of the joint system. Where do these states come from? What we are always meaning in a situation such as this is that the system under consideration (here, the ancilla) has been measured in a basis of which the relevant state is a part, and then one basis state over others `picked out'. Physically, it is picked out by being the state which is relative to some other state that we are interested in. At some point, then, in a tractable analysis we will have to simply stop and pick out the state relative to which we wish to find other states by fiat. At this point we construct the operators for this `ultimate' state from the density operator that we wish it to have, and then find everything else relative to it. Where this operation takes place is entirely a matter of convenience. What must be remembered, though, is that at this cut-off point we then lose the record of the interactions of that system with other systems. For example, we could not in the current situation choose to construct the operators for the ancilla from the ground up -- that would give
$$  \mathbf{q}_2 = \left( \begin{array}{c}\openone \otimes \sigma_x\\ \pm\openone \otimes \sigma_y \\ \pm\openone \otimes \sigma_z \end{array} \right)$$
 
\noindent comparing this with (\ref{psapihy}), we see that we have lost the information about the first system that was contained in the ancilla.\\

In this situation, what we need to do is push the ultimate state back one system further, and have the ancilla measured by a third system. If we start that system in the $\ket{\mathbf{0}}$ and perform CNOT with the ancilla as control and the third system as target then we have 
$$ \mathbf{q}_2 = \left( \begin{array}{c}\openone \otimes \sigma_x \otimes \sigma_x \\ \sigma_x \otimes \sigma_y \otimes \sigma_x \\ \sigma_x \otimes \sigma_z\otimes \openone \end{array} \right) \ \ \ \ \ \ \ \ \ \ \ \mathbf{q}_3 = \left( \begin{array}{c}\openone \otimes \openone \otimes \sigma_x \\ \sigma_x \otimes \sigma_z  \otimes \sigma_y \\ \sigma_x \otimes \sigma_z \otimes \sigma_z \end{array} \right)$$

\noindent The operators of the ancilla relative to the states \ket{0} and \ket{1} of the third system are, respectively, $\mathbf{q}^+_2$ and $\mathbf{q}^-_2$:
\begin{eqnarray} \mathbf{q}^{\pm}_2 & = & \mathbf{q}_2 ( \openone \pm q_{3z} )\nonumber\\
{} & = & \left( \begin{array}{c} (\openone \otimes \sigma_x \pm \sigma_x \otimes \sigma_z ) \otimes (\sigma_x \pm  \sigma_z)\\
 (\sigma_x \otimes \sigma_y \pm \sigma_x \otimes \sigma_z ) \otimes (\sigma_x \pm  \sigma_z) \\
 (\sigma_x \otimes \sigma_z \pm \sigma_x \otimes \sigma_z) \otimes (\sigma_x \pm \sigma_z)\end{array} \right) \label{lafa} \end{eqnarray}

\noindent These then are the operators for the ancilla to be in the states \ketbra{0}{0} and \ketbra{1}{1}, relative to which the first system states are (\ref{prone}) and (\ref{prtwo}).\\

If we look at (\ref{prone}), (\ref{prtwo}) and (\ref{lafa}) we notice something interesting: the sum of the relative $q_{ai}$ is (neglecting a factor of 2) the original $q_{ai}$ for the system. This is, in fact, the case for any size system where the states relative to it are a complete POVM -- the sum of the $q_{ai}$ relative to each of the states of the POVM is the $q_{ai}$ for the system before measurement. The proof is straightforward. Using (\ref{jfda}) we have
$$ \sum_{i=1}^m \mathbf{q}_1^{(i)}  =  \mathbf{q}_1 (m\openone + \sum_n \sum_i \av{\sigma_n}_{\ketbra{i}{i}} q_{2n} ) $$

\noindent Now
\begin{eqnarray*} \sum_a \bra{a} A \ket{a} & = & \sum_a \trace (A\ketbra{a}{a}) \\
{} & = & \trace \left( A \sum_a \ketbra{a}{a} \right)\end{eqnarray*}

\noindent If $\{ \ketbra{a}{a} \}$ forms a POVM then $\sum_a \ketbra{a}{a} = \openone$, so
$$ \sum_a \bra{a} A \ket{a} = \trace A $$

\noindent Therefore, as the $\sigma_n$ have zero trace, we have
\begin{eqnarray*} \sum_{i=1}^m \mathbf{q}_1^{(i)}  & = &  \mathbf{q}_1 (m\openone + \sum_n \trace (\sigma_n) q_{2n} )\\
{} & = & m \mathbf{q}_1 \end{eqnarray*}

Physically, then, we have performed a POVM measurement on the ancilla system, and are now looking at states of the first system with respect to elements of the POVM.

\section{Mixed states}\label{ddhrms}

So far we have been considering only the representation of pure states in the Deutsch-Hayden picture. We shall see that one of the advantages of this picture is that the representation of mixed states does not differ in kind from the pure-state representations. We will also see that the Deutsch-Hayden representation of mixed states is transparent to their physical origins.

The Deutsch-Hayden operators $q_{ai}$ form a basis in the Hilbert-Schmidt space. This is not, however, an arbitrary basis: the elements of the individual $\mathbf{q}_a$ combine to form the elements of the basis. We have, in fact, a \emph{product} basis on the Hilbert space. We can therefore consider the space of the system as a product Hilbert space, of factor spaces both of which (or each of which for more than two systems) are co-ordinatised by $\mathbf{q}_a$. These factor spaces will, of course, evolve over time -- the product does not remain fixed. However, at each time $t$ the space of the system may be written as the product $\h{1}(t) \otimes \h{2}(t)$ where $\{ q_{1i}(t)\}$ forms a basis on $\h{1}$ and $\{ q_{2j}(t)\}$ forms one on $\h{2}$.

In order to see to what these factor spaces correspond, consider the situation at time $t=0$. In this case $\mathbf{q}_1 = \mathbf{\sigma} \otimes \openone$ and $\mathbf{q}_2 = \openone \otimes \mathbf{\sigma}$, and in the Schr\"{o}dinger representation we have the state \ket{00}. The spaces co-ordinatised by the $\mathbf{q}_a$ are therefore transparently the spaces corresponding to operators of the two systems. Let us look at, for example, an operator that operates only on the first system: $\hat{A}\otimes \openone$. This can be written
\begin{eqnarray*} \hat{A}\otimes \openone & = & \sum_i a_i \sigma_i \otimes \openone \\
{} & = & \sum_i a_i q_{1i}(0) \end{eqnarray*}

\noindent Therefore at a given subsequent time $t$ we have
$$ \hat{A}(t) = \sum_i a_i q_{1i}(t) $$

\noindent That is, anything that can be said about the first system alone is contained within the $q_{1i}$ alone, with no reference to other \q. The space co ordinatised by $q_{1i}(t)$ at any given time is therefore the space of operators on the first system alone. Thus, at any given time, the overall space of the system is a product space of the individual system spaces. Unlike in a standard Hilbert space representation, this is the case always, regardless of whether the systems are entangled or not.

In none of the foregoing have we had to say whether the $q_{ai}$ correspond to pure or to mixed states. It might be questioned whether mixed states can indeed be accommodated in this picture -- after all, we have constructed $q_{ai}$ by evolving them unitarily from a fixed state, and we know that unitary evolution alone cannot produce mixed states from pure.

The key to representing mixed states is the fact that all mixed states can be considered as pure states where one or more systems have been traced out \cite[pp110ff]{nandc}. The Deutsch-Hayden operators for a mixed state on a system $a$ will be the $q_{ai}$ which have been evolved as part of a larger, pure system. For example, we have found the $q_{ai}$ corresponding to \ket{00} + \ket{11} on two systems. If we trace out, for example, the second system then we are left with the maximally mixed state \ketbra{0}{0} + \ketbra{1}{1}. This is represented in Deutsch-Hayden terms by
\begin{equation} \mathbf{q}_1 = \left( \begin{array}{c}\sigma_z \otimes \sigma_x \\ -\sigma_y \otimes \sigma_x \\ \sigma_x \otimes \openone \end{array} \right) \label{simplemixed} \end{equation}

We can see that one of the most immediate differences between pure and mixed states is the dimension of the non-trivial part of the subspace for the system. A pure state of the first system would be represented by operators of the form $\sigma_i \otimes \openone$ -- for example, \ket{0} + \ket{1} is given by
$$ \mathbf{q}_1 = \left( \begin{array}{c}\sigma_z \otimes \openone \\ -\sigma_y \otimes \openone \\ \sigma_x \otimes \openone \end{array} \right)$$

\noindent by contrast, the subspace of a mixed-state system is non-trivial in more dimensions, corresponding to the systems which have been traced over to get from the pure to the mixed state. Note that in the Deutsch-Hayden picture there is no such thing as tracing out a system: whether or not we are looking at the second system makes no difference to the way in which operators on the first system are constructed.

It is a necessary and sufficient condition of a state being mixed that its $q_{ai}$ cannot be written non-trivially on the same size subspace that a pure state of the same system would require. If this were possible, then the operators in the non-trivial part of the subspace could be written
$$ \mathbf{q} = U^\dagger \mathbf{\sigma}\otimes \mathbf{\sigma}\otimes \ldots \otimes \mathbf{\sigma} U$$

\noindent for some unitary operator $U$. The average values of the $\mathbf{q}$ could then be written
\begin{eqnarray*} \av{\mathbf{q}} & = & \trace ( \mathbf{q} \ketbra{\mathbf{0}}{\mathbf{0}} ) \\
{} & = & \trace (U^\dagger \mathbf{\sigma}\otimes \mathbf{\sigma}\otimes \ldots \otimes \mathbf{\sigma} U \ \ketbra{\mathbf{0}}{\mathbf{0}} ) \\
{} & = & \trace (\mathbf{\sigma}\otimes \mathbf{\sigma}\otimes \ldots \otimes \mathbf{\sigma} \ U \ketbra{\mathbf{0}}{\mathbf{0}} U^\dagger )\\
{} & = & \trace (\mathbf{\sigma}\otimes \mathbf{\sigma}\otimes \ldots \otimes \mathbf{\sigma} \ \rho ) \end{eqnarray*}

\noindent Note that $\rho$ in the final line is pure, which gives the contradiction: these are the average values of a basis set in some pure state, which cannot give the same outcomes as a mixed state. We will therefore always need a larger subspace to describe a mixed system than a pure one.

As well as distinguishing pure and mixed states in this way, we can also frame the standard condition $\trace \rho^2 < 1$ in terms of the $q_{ai}$ (here for 2 systems):
\begin{eqnarray*} \rho_{12} & = & \frac{1}{4} \Big( \sum_{ij} a_{ij} \sigma_i \otimes \sigma_j \Big) \\
\rho^2 & = & \frac{1}{16} \Big( \sum_{ijmn} a_{ij} a_{mn} \sigma_i \sigma_m \otimes \sigma_j\sigma_n\Big) \\
\trace \rho^2 & = & \frac{1}{16} \Big( a_{00}^2 \trace (\openone \otimes \openone ) +  \sum_{jn} a_{0j} a_{0n}\trace (\openone \otimes \sigma_j \sigma_n) \\
{} & {} & {}  + \sum_{im} a_{i0} a_{m0} \trace (\sigma_i \sigma_m \otimes \openone) + \sum_{ij} a_{ij}^2 \trace (\sigma_i^2 \otimes \sigma_j^2) \Big) \\
{} & = & \frac{1}{16} \Big( 4 + 4( \sum_{ij} a_{i0}^2 + a_{0j}^2 + a_{ij}^2 )\Big)\\
{} & = & \frac{1}{4} \Big( 1 + \sum_{ij=1}^3 \av{q_{1i}}^2 + \av{q_{2j}}^2 + \av{q_{1i}q_{2j}}^2 \Big) \end{eqnarray*}

\noindent The condition for a mixed state is therefore
\begin{equation} \sum_{ij=1}^3 \av{q_{1i}}^2 + \av{q_{2j}}^2 + \av{q_{1i}q_{2j}}^2 < 3 \label{cjka}\end{equation}

The Schmidt decomposition condition for pure states can also be framed in terms of the \q. The Schmidt decomposition of a pure state density matrix is
$$ \rho_{AB} = \sum_{ij} a_{ij} \ketbra{i}{j} \otimes \ketbra{i}{j} $$

\noindent Writing this in terms of $\sigma$-matrices, that is
\begin{eqnarray*}  \rho_{AB}  & = &  a (\openone + \sigma_z)\otimes (\openone + \sigma_z) + b ( \sigma_x - \mi \sigma_y ) \otimes ( \sigma_x - \mi \sigma_y ) \\
& & {} + c(\sigma_x + \mi \sigma_y ) \otimes (\sigma_x + \mi \sigma_y )   + d (\openone - \sigma_z) \otimes (\openone - \sigma_z) \end{eqnarray*}

\noindent The coefficients are therefore (in terms of the \q)
\begin{eqnarray*} \frac{a}{2} & = & \av{\openone \otimes \openone} + \av{q_{1z}} + \av{q_{2z}} + \av{q_{1z}q_{2z}} \\
\frac{b}{2} & = & \av{q_{1x}q_{2x}} - \av{\mi q_{1y}q_{2x}} -\av{\mi q_{1x}q_{2y}} - \av{q_{1y}q_{2y}} \\
\frac{c}{2} & = & \av{q_{1x}q_{2x}} + \av{\mi q_{1y}q_{2x}} +\av{\mi q_{1x}q_{2y}} - \av{q_{1y}q_{2y}}\\
\frac{d}{2} & = & \av{\openone \otimes \openone} - \av{q_{1z}} - \av{q_{2z}} + \av{q_{1z}q_{2z}} \end{eqnarray*}

\noindent These follow the usual Schmidt rule $a^2 + b^2 + c^2 + d^2 = 1$, which when expanded in these terms gives (\ref{cjka}).

As well as distinguishing mixed states when we are presented with them, we can also construct the $q_{ai}$ for mixed states in the same way that we constructed them for pure states. We can either construct the purified state using a circuit, or we can directly construct from the density operator. Let us look at both of these in turn for the simple states \ketbra{0}{0} + \ketbra{1}{1} given by (\ref{simplemixed}).

Firstly we will try the purification method. This is straightforward \cite[pp110ff]{nandc}: take the mixed density operator in its diagonal basis and then add another system with the same basis, to make a pure state whose Schmidt decomposition for the original system gives the basis in which the density operator is diagonal. In the case of \ketbra{0}{0} + \ketbra{1}{1} we add a second system with basis $\{ 0,1 \} $. Possible purifications are \ket{00} + \ket{11} and \ket{01} + \ket{10}. We now find the corresponding $q_{ai}$ for the whole system, and then take only the $q_{ai}$ for the original system. In the first case we would have (\ref{simplemixed}). The second can be found by flipping the second bit in a state \ket{00} + \ket{11}, which has the effect on the $q_{ai}$ of being acted on by $\openone \otimes \sigma_x$. As it happens, this gives us back the same \q, (\ref{simplemixed}).

Let us now try direct construction. We have the density operator for the state,
$$\rho = \twtw{1}{0}{0}{1}$$
\noindent which tells us
$$ \av{q_i} = 0$$

\noindent We know that we will not be able to express the $q_i$ as combinations of $\sigma$ (in this case as well this is obvious), so we will add an extra two dimensions to the subspace under consideration and try combinations of $\sigma_i \otimes \sigma_j$. We have quite a lot more freedom in this construction than in that of the pure states, as we do not care what the operators for the ancillary system are. All we need to make sure is that the operators for the system we do care about co-ordinatise a well-formed subspace of space for the whole system. It is easiest to start off with $\mathbf{\sigma}$ in the first position, and then add $\sigma_x$ where needed, giving
$$ \mathbf{q}_1 = \left( \begin{array}{c}\sigma_x \otimes \openone \\ \sigma_y \otimes \sigma_x \\ \sigma_z \otimes \sigma_x \end{array} \right) $$

\noindent We verify that this is indeed a representation of \ketbra{0}{0} + \ketbra{1}{1} as they are the $\mathbf{q}_1$ operators from the set $q_{ai}^{(6)}$ in the previous section.

This representation of mixed states is a very physical one. The Deutsch-Hayden picture does not recognise any form of evolution other than unitary evolution, and this is reflected in the fact that even when a set of $q_{ai}$ for a mixed state are directly constructed as above, their structure shows that they are in fact part of a larger pure state. The size of the mixed-state operators can be found from the purification method used above. An extra system is needed to purify each system that is mixed -- so the Deutsch-Hayden operators will in general be twice as long for mixed states than for pure states of the same sized system. That is, a mixed state of $N$ qubits will in general be part of a pure state of dimension $2.2^N$, and the Deutsch-Hayden descriptors will have length $2N$.

\section{Reduced descriptors}\label{mixdsts}


We have seen that an $n$-dimensional mixed system needs to be written on a larger space than a corresponding pure system. We will now look in detail at the ability to write a descriptor on a reduced space, in particular with reference to entanglement between systems.

Let us take a three qubit system, the overall state of which is pure, and then neglect the third qubit (that is, we are only interested in the descriptors for the first two systems). If the first two systems together are not entangled with the third then their descriptors will satisfy
\begin{eqnarray*} \rho_{123} & = & \rho_{12}\rho_3 \\
\av{\mathbf{q}_{123}} \mathbf{\sigma} \otimes \mathbf{\sigma} \otimes \mathbf{\sigma} & = & \av{\mathbf{q}_{12}} \mathbf{\sigma} \otimes \mathbf{\sigma} \ \av{\mathbf{q}_3} \mathbf{\sigma}\\
\av{\mathbf{q}_1 \mathbf{q}_2 \mathbf{q}_3} & = & \av{\mathbf{q}_1 \mathbf{q}_2}\av{\mathbf{q}_3} \end{eqnarray*}

\noindent Writing this out in full, with $U$ giving the evolution of all three systems gives us
\begin{eqnarray*} \lefteqn{\trace ( \ketbra{\mathbf{0}}{\mathbf{0}}  U^{\dagger} \mathbf{\sigma} \otimes \mathbf{\sigma} \otimes \mathbf{\sigma} U )}\\
& = & \trace ( \ketbra{\mathbf{0}}{\mathbf{0}}  U^\dagger \mathbf{\sigma} \otimes \mathbf{\sigma} \otimes \openone U ) \trace (  \ketbra{\mathbf{0}}{\mathbf{0}} U^\dagger \openone \otimes \openone \otimes \mathbf{\sigma} U )\\
\lefteqn{ \trace ( \mathbf{\sigma} \otimes \mathbf{\sigma} \otimes \mathbf{\sigma} \ U^\dagger \ketbra{\mathbf{0}}{\mathbf{0}}  U )}\\
& = & \trace ( \ketbra{\mathbf{0}}{\mathbf{0}} U^\dagger \mathbf{\sigma} \otimes \mathbf{\sigma} \otimes \openone U ) \trace (  \openone \otimes \openone \otimes \mathbf{\sigma} \ U^\dagger \ketbra{\mathbf{0}}{\mathbf{0}}U)\end{eqnarray*}

\noindent Writing $\mathbf{a} \otimes \mathbf{b} \otimes \mathbf{c} \equiv U^\dagger \ketbra{\mathbf{0}}{\mathbf{0}} U$ we have
\begin{eqnarray*} \lefteqn{ \trace ( \mathbf{\sigma} \otimes \mathbf{\sigma} \otimes \mathbf{\sigma} \ \mathbf{a} \otimes \mathbf{b} \otimes \mathbf{c} )}\\
& = & \trace( \ketbra{\mathbf{0}}{\mathbf{0}} \ U^\dagger \mathbf{\sigma} \otimes \mathbf{\sigma} \otimes \openone U ) \ \trace (\openone \otimes \openone \otimes \mathbf{\sigma} \ \mathbf{a} \otimes \mathbf{b} \otimes \mathbf{c} ) \\
{} & = & \trace( \ketbra{\mathbf{0}}{\mathbf{0}} \ U^\dagger \mathbf{\sigma} \otimes \mathbf{\sigma} \otimes \openone U ) \ \trace ( \mathbf{a} \otimes \mathbf{b} \otimes \mathbf{\sigma c} ) \\
\lefteqn{ \trace (\mathbf{\sigma} \otimes \mathbf{\sigma} \otimes \openone \ \mathbf{a} \otimes \mathbf{b} \otimes \openone ) \ \trace (\mathbf{\sigma c} )}\\
& = & \trace ( \ketbra{\mathbf{0}}{\mathbf{0}} \ U^\dagger \mathbf{\sigma} \otimes \mathbf{\sigma} \otimes \openone U ) \ \trace (\mathbf{a} \otimes \mathbf{b} \otimes \openone ) \trace (\mathbf{\sigma c} ) \\
\lefteqn{\trace (\mathbf{\sigma} \otimes \mathbf{\sigma} \otimes \openone \ \mathbf{a} \otimes \mathbf{b} \otimes \openone ) }\\
& = & \trace ( \ketbra{\mathbf{0}}{\mathbf{0}} \ U^\dagger \mathbf{\sigma} \otimes \mathbf{\sigma} \otimes \openone U ) \ \trace (\mathbf{a} \otimes \mathbf{b} \otimes \openone )\end{eqnarray*}

\noindent Now in general $\mathbf{a}$, $\mathbf{b}$ and $\mathbf{c}$ are sums, $\sum_{i=0}^3 t_i \sigma_i$, where $t_0^{(a)}t_0^{(b)}t_0^{(c)}$ is the $\openone\otimes \openone$ component of $\mathbf{a} \otimes \mathbf{b} \otimes \mathbf{c}$. As this component is never zero, $t_0^{(a,b,c)} \neq 0$. Therefore the trace of $\mathbf{a}$, $\mathbf{b}$ and $\mathbf{c}$ will always be 1 (neglecting normalisation) - and hence $\trace (\mathbf{a} \otimes \mathbf{b} \otimes \openone )$ is a simple, constant numerical factor which can be neglected here, and can be incorporated into the normalisation of the descriptors in the end. We therefore have
\begin{eqnarray*} \trace ( \ketbra{\mathbf{0}}{\mathbf{0}} \ U^\dagger \mathbf{\sigma} \otimes \mathbf{\sigma} \otimes \openone U ) & = & \trace (\mathbf{\sigma} \otimes \mathbf{\sigma} \otimes \openone \ \mathbf{a} \otimes \mathbf{b} \otimes \openone )\\
\av{\mathbf{q}_1 \mathbf{q}_2} & = & \trace (\mathbf{\sigma} \otimes \mathbf{\sigma} \ \mathbf{a} \otimes \mathbf{b})\end{eqnarray*}

\noindent Now let us define $U_{12}$ as the elements of the matrix $U$ that act only on $\h{12}$. We then have
\begin{eqnarray}\av{\mathbf{q}_1 \mathbf{q}_2} & = & \trace ( \mathbf{\sigma} \otimes \mathbf{\sigma} \ U_{12}^\dagger\ketbra{\mathbf{0}}{\mathbf{0}}U_{12})\nonumber\\
\av{\mathbf{q}_1 \mathbf{q}_2} & = & \trace (U_{12}^\dagger \mathbf{\sigma} \otimes \mathbf{\sigma} U_{12}\ \ketbra{\mathbf{0}}{\mathbf{0}})\label{apa}\end{eqnarray}

\noindent What then are the expressions $U_{12}^\dagger \mathbf{\sigma} \otimes \mathbf{\sigma} U_{12}$ exactly? They are the same as the descriptors $U^\dagger \mathbf{\sigma} \otimes \mathbf{\sigma} \otimes \openone U$ where only the action on $\h{12}$ is taken into account. We can therefore see that they are the ``simply reduced'' $\mathbf{q}_1 \mathbf{q}_2$; that is, the descriptors with only their $\h{12}$ components. For example, if $\mathbf{q}_1 = \sigma_x \otimes \sigma_z \otimes \sigma_y$ then the simply reduced form on $\h{12}$ is $\sigma_x \otimes \sigma_z$.

\noindent At this point we need to introduce some terminology. We will say that:
\begin{quotation} A descriptor $\mathbf{q}$ is said to be \textbf{represented by} an expression $\mathbf{n}$ when it is the case that $ \av{\mathbf{q}} = \av{\mathbf{n}}$ \end{quotation}

\noindent We saw examples of representation previously, where many different sets of $\mathbf{q}$'s described the same density matrix; each set was represented by the others. However, representation is not restricted to such examples on the same Hilbert space: in general the dimensionality of $\mathbf{n}$ and $\mathbf{q}$ differ.

Looking at (\ref{apa}) we can see that this is a case of representation. We have shown that $\mathbf{q}_1\mathbf{q}_2 \in \h{123}$ can be represented by $ U_{12}^\dagger \mathbf{\sigma} \otimes \mathbf{\sigma} U_{12} \in \h{12}$. This is therefore the key to representing pure states. When looking at a system $s$, only the elements of the descriptor on $\h{s}$ need be used. When considering only that system, the components on other spaces make no difference to the average values, and might as well be replaced by $\openone$. 

As a consequence, when looking at the past evolution of the system we may as well replace the action of $U$ on $\h{3}$ by $\openone$ as well.

(It becomes necessary here to introduce some notation. There is the potential for confusion between descriptors with different numbers of components. When the context does not make clear on which space they have support then we will write for example $[ \mathbf{q}_1]_{12}$ for the descriptor $\mathbf{q}_1$ with components only on $\h{12}$ -- that is, $U_{12}^\dagger \mathbf{\sigma} \otimes \openone U_{12}$.) 

Another way of seeing this is by considering the following way of expressing the descriptors for the first two systems:
\begin{eqnarray} \av{\mathbf{q}_1\mathbf{q}_2} & = & \trace ( \ketbra{\mathbf{0}}{\mathbf{0}} \ U^\dagger \mathbf{\sigma}\otimes \mathbf{\sigma}\otimes \openone U ) \nonumber \\
{} & = &  \trace ( U^\dagger\ketbra{\mathbf{0}}{\mathbf{0}} U\  \mathbf{\sigma}\otimes \mathbf{\sigma}\otimes \openone )\nonumber\\
{} & = & \trace ( U_{12}^\dagger \ketbra{00}{00} U_{12} \ \mathbf{\sigma}\otimes \mathbf{\sigma} ) \trace ( U_3^\dagger \ketbra{0}{0}U) \nonumber\\
{} & = & \av{[ \mathbf{q}_1 \mathbf{q}_2]_{12}} \ \trace ( U_3^\dagger \ketbra{0}{0}U_3) \label{vdios}\end{eqnarray}

\noindent So what happens when there is no entanglement with qubit 3 is that the expression $\trace ( U_3^\dagger \ketbra{0}{0}U_3)$ becomes irrelevant; it is not zero when $\av{[ \mathbf{q}_1 \mathbf{q}_2]_{12}}$ is not zero. 

We have seen what happens when there is no entanglement between the two systems. What changes then if the first two qubits are in a \emph{mixed} state -- if they are entangled with the third qubit? If we restrict ourselves to simply that system, what can we say about the qubits and their descriptors? The most obvious point is that we will no longer be able to describe them on $\h{12}$ using the simply-reduced descriptors. The information contained in $\h{3}$ becomes relevant to describing the past evolution of the two-qubit system -- as shown in Chapter 2, we require $\h{123}$ to write them on.

We now have a problem. As can be seen from all the foregoing work on the Deutsch-Hayden approach there is no operation in the formalism that corresponds to the action of tracing out a system, which is the usual way of getting a mixed state representation. Subsystems and their descriptors have been considered in their physical context, as part of the wider system, and their descriptors written on the space of that wider system. The question is: given that there is no \emph{physical} operation corresponding to tracing out a system, what happens when we have no knowledge of anything other than the system under consideration? The solution is prompted by the foregoing paragraph: that ignoring the action of the third system means ignoring $\h{3}$. So if we want to restrict our attention simply to the system of qubits 1 and 2 then we can only have access to $\h{12}$ on which to write our descriptors. Hence the problem.

This may appear to be labouring the point somewhat as it is obvious what the solution to this is going to be. It is, however, instructive to show exactly how that comes about in the Deutsch-Hayden approach, and why it is needed -- why we cannot just deal with the descriptors as we have them on $\h{123}$.

The solution is that we have found a set of operators on $\h{123}$ which we know can be simply reduced to $\h{12}$. If we can write our `mixed' descriptors in terms of these `pure' descriptors then we have a way of representing the descriptors associated with mixed states on $\h{12}$. There are two ways in which we could do this. Either a combination of the descriptors associated with pure states could represent the descriptor we are concerned with, or else the pure descriptors could form a complete basis on $\h{123}$ for all possible $u^\dagger \mathbf{\sigma} \otimes \mathbf{\sigma} \otimes \openone U$.
 
We know from standard quantum mechanics that the first is going to be the case -- this is the standard way of writing mixed states as a mixture of pure ones. We can show this in the present case. Let us look again at (\ref{vdios}). We can re-write this as
$$ \av{\mathbf{q}_1\mathbf{q}_2} = \trace \Big( \ketbra{00}{00} \ ( U_{12}^\dagger \mathbf{\sigma} \otimes \mathbf{\sigma} U_{12} \trace ( U_3^\dagger \ketbra{0}{0}U_3 ) )  \Big) $$

\noindent Writing
$$ \mathbf{n}\otimes \mathbf{m} = U_{12}^\dagger \mathbf{\sigma} \otimes \mathbf{\sigma} U_{12} \trace ( U_3^\dagger \ketbra{0}{0}U_3 ) $$

\noindent we have
$$ \av{\mathbf{q}_1\mathbf{q}_2} = \trace ( \ketbra{00}{00} \ \av{\mathbf{n}\otimes \mathbf{m}}) $$

\noindent That is, we can represent $\mathbf{q}_1\mathbf{q}_2$ on $\h{12}$, by the expression $\mathbf{n}\otimes\mathbf{m}$. Now, $\mathbf{n}\otimes\mathbf{m}$ can be expressed in terms of a complete basis $\{\mathbf{\lambda}_i\} \in \h{12}$:
$$ \av{\mathbf{q}_1\mathbf{q}_2} = \trace ( \ketbra{00}{00} \ \av{\sum_i \lambda_i \mathbf{\lambda}_i}) $$

\noindent We know one candidate for $\{\lambda_i \}$: the set of all possible descriptors on $\h{12}$, namely $U^\dagger \mathbf{\sigma}\otimes \mathbf{\sigma}U$. We can therefore write
\begin{eqnarray} \av{\mathbf{q}_1\mathbf{q}_2} & = & \trace \left( \ketbra{\mathbf{0}}{\mathbf{0}} \ \sum_i w_i U_i^\dagger \mathbf{\sigma}\otimes \mathbf{\sigma}U_i \right)\nonumber\\
{} & = & \trace \left( \ketbra{\mathbf{0}}{\mathbf{0}} \ \sum_i w_i U_i^\dagger \mathbf{\sigma}\otimes \mathbf{\sigma}\otimes \openone U_i \right)\label{sasnv}\\
{} & = & \sum_i w_i \av{\mathbf{q}_{1i}\mathbf{q}_{2i}} \label{kjsla} \end{eqnarray}

\noindent where 
$$ w_i = \trace ( U_i^\dagger \mathbf{\sigma} \otimes \mathbf{\sigma} U_i \ U_{12}^\dagger \mathbf{\sigma}\otimes\mathbf{\sigma} U_{12} ) \ \trace ( U_3^\dagger \ketbra{0}{0} U_3 )$$

\noindent and where the $\mathbf{q}_{ni}$ (given by the $U_i$) are those descriptors on $\h{123}$ that describe qubits 1 and 2, unentangled with qubit 3.

What about the second possibility, that any given $\mathbf{q}_1 \mathbf{q}_2 \in \h{123}$ can be written as a mixture of the pure $\mathbf{q}_{1i} \mathbf{q}_{2i}$, rather than simply being represented by them? This would require us to be able to write
\begin{equation} \mathbf{q}_1\mathbf{q}_2 = \sum_i \lambda_i U_i^\dagger \mathbf{\sigma} \otimes \mathbf{\sigma} \otimes \openone U_i \label{oifh} \end{equation}

\noindent where again the $U_i$ give qubits 1 and 2 unentangled with qubit 3. We have written $\lambda_i$ here rather than $w_i$ as they do not have the same form:
$$ \lambda_i = \trace ( U_i^\dagger \mathbf{\sigma} \otimes \mathbf{\sigma} \otimes \openone U_i \ U_{123}^\dagger \mathbf{\sigma} \otimes \mathbf{\sigma} \otimes \openone U_{123} )$$

\noindent where $U_{123}$ gives the first two qubits in a mixed state.

We can show if this is the correct form of $\mathbf{q}_1\mathbf{q}_2$ by considering the physical predictions that it would give. We know that the correct values for the averages of these descriptors are given by (\ref{kjsla}). Writing $\mathbf{q}_1\mathbf{q}_2$ as in (\ref{oifh}) would give as the average values
$$\av{\mathbf{q}_1\mathbf{q}_2} = \trace \left( \ketbra{\mathbf{0}}{\mathbf{0}} \ \sum_i \lambda_i U_i^\dagger \mathbf{\sigma}\otimes \mathbf{\sigma}\otimes \openone U_i \right)$$

\noindent Comparing this with (\ref{sasnv}) we see that these can only be identical if $\lambda_i = w_i$. Suppose that this is the case. We would then have
\begin{eqnarray*} \lefteqn{ \trace ( U_i^\dagger \mathbf{\sigma} \otimes \mathbf{\sigma} U_i \ U_{12}^\dagger \mathbf{\sigma}\otimes\mathbf{\sigma} U_{12} ) \ \trace ( U_3^\dagger \ketbra{0}{0} U_3 ) } \\
& & = \trace ( U_i^\dagger \mathbf{\sigma} \otimes \mathbf{\sigma} \otimes \openone U_i \ U_{123}^\dagger \mathbf{\sigma} \otimes \mathbf{\sigma} \otimes \openone U_{123} )
\end{eqnarray*}

\noindent Now let us write
\begin{eqnarray*} U_i^\dagger \mathbf{\sigma} \otimes \mathbf{\sigma} \otimes \openone U_i & = & U_i^\dagger \mathbf{\sigma} \otimes \mathbf{\sigma} U_i \otimes \mathbf{\gamma}_i\\
U_{123}^\dagger \mathbf{\sigma} \otimes \mathbf{\sigma} \otimes \openone U_{123} & = &  U_{12}^\dagger \mathbf{\sigma}\otimes\mathbf{\sigma} U_{12} \otimes \mathbf{c} \end{eqnarray*}

\noindent This gives us
\begin{eqnarray*} \lefteqn{ \trace ( U_i^\dagger \mathbf{\sigma} \otimes \mathbf{\sigma} U_i \ U_{12}^\dagger \mathbf{\sigma}\otimes\mathbf{\sigma} U_{12} ) \ \trace ( U_3^\dagger \ketbra{0}{0} U_3 )} \\
& & = \trace ( U_i^\dagger \mathbf{\sigma} \otimes \mathbf{\sigma} U_i \ U_{12}^\dagger \mathbf{\sigma}\otimes\mathbf{\sigma} U_{12} ) \ \trace ( \mathbf{\gamma}_i \mathbf{c} )
\end{eqnarray*}

\noindent which would require that for all $i$ such that $\trace ( U_i^\dagger \mathbf{\sigma} \otimes \mathbf{\sigma} U_i \ U_{12}^\dagger \mathbf{\sigma}\otimes\mathbf{\sigma} U_{12} ) \neq 0$,
$$ \trace ( U_3^\dagger \ketbra{0}{0} U_3 ) = \trace ( \mathbf{\gamma}_i \mathbf{c} )$$

\noindent which is a contradiction, as the left hand side is a constant value for all $i$ whereas the right hand side is a variable.

It is therefore not the case that $\mathbf{q}_1\mathbf{q}_2$ itself can be written as a mixture of descriptors for pure states, but it is the case that it can be \emph{represented} by such.

This is a very interesting situation. What this shows is that in the Deutsch-Hayden approach the term ``mixture'' is fundamentally incorrect. The descriptors for such systems can be represented by a mixture of descriptors for pure states, but they are not themselves identical with a mixture, even when looking at the pure descriptors spanning the original Hilbert space. All of this is very different from a standard density-matrix formalism, where the pure state density matrices co-ordinatise the space of possible density matrices and are hence `special' in a way that the mixed density matrices are not. Indeed, the mixed density matrices are given as different types of entities from the pure state ones (a fact emphasised by the standard Schr\"{o}dinger notation, which cannot deal with them).

In terms of the Deutsch-Hayden approach, mixed and pure states are both described by sets of descriptors on a Hilbert space of certain dimensions, this corresponding to the number of systems involved in the evolution of the system under consideration. The difference between pure and mixed comes when we wish to reduce the Hilbert space on which we write our descriptors -- that is, when we wish to ignore certain subsystems in the analogous operation to tracing out systems. Only then does any notational difference arise, some systems needing a sum of descriptors to describe them, some able to use the reduced descriptors. However, they are still both on a par with each other: the descriptors corresponding to pure states are not `special' in any fundamental way, and they do not co-ordinatise the space of all descriptors. Each are as fundamental as the other, without the definition of one being dependent on the other.

\section{Entanglement swapping}

We will now look at the flow information in an entanglement swapping situation. In the Schr\"{o}dinger representation, the particular example that we will use is the following \cite{entswap}. We have four qubits in the overall state $\ket{\Phi}$:
\begin{eqnarray*}
\ket{\Phi} & = & (\ket{00} + \ket{11})_{12} \otimes (\ket{00} + \ket{11})_{34}\\
& = & \ket{00}_{23}\ket{00}_{14} + \ket{01}_{23}\ket{01}_{14} + \ket{10}_{23}\ket{10}_{14} + \ket{11}_{23}\ket{11}_{14}\\
& = & (\ket{00} + \ket{11})_{23} \ \otimes \ (\ket{00} + \ket{11})_{14}\\
& & {}+ (\ket{00} - \ket{11})_{23} \ \otimes \ (\ket{00} - \ket{11})_{14}\\
& & {}+ (\ket{01} + \ket{10})_{23} \ \otimes \ (\ket{01} + \ket{10})_{14}\\
& & {}+ (\ket{01} - \ket{10})_{23} \ \otimes \ (\ket{01} - \ket{10})_{14}\end{eqnarray*}

\noindent If we then measure qubits 2 and 3 in the Bell basis, we can see that qubits 1 and 4 become entangled, and the measurement destroys the entanglement within the pairs (1,2) and (3,4) -- it has ``swapped" to the pair (1,4).

The locality issues of this operation are well known: the qubit pairs (1,4) and (2,3) could be sent to opposite ends of the galaxy after the original entangling operations, and yet a subsequent operation on (2,3) will entangle (1,4). Even worse, qubits 1 and 4 could be separated before the measurement on (2,3), yet after it they are in a maximally entangled state, which can for example be used to teleport between them.

We start our analysis with the four qubits in the state $\ket{\Phi}$:
\begin{eqnarray*} \mathbf{q}_1 = \left( \begin{array}{c} \sigma_z \otimes \sigma_x \otimes \openone \otimes \openone\\  -\sigma_y \otimes \sigma_x \otimes \openone \otimes \openone \\ \sigma_x \otimes \openone \otimes \openone \otimes \openone \end{array} \right) & {}\ \ \ & \mathbf{q}_2 = \left( \begin{array}{c} \openone \otimes \sigma_x \otimes \openone \otimes \openone\\  \sigma_x \otimes \sigma_y \otimes \openone \otimes \openone \\ \sigma_x \otimes \sigma_z \otimes \openone \otimes \openone \end{array} \right) \\
\mathbf{q}_3 = \left( \begin{array}{c} \openone \otimes \openone \otimes \sigma_z \otimes \sigma_x\\  - \openone \otimes \openone \otimes \sigma_y \otimes \sigma_x \\ \openone \otimes \openone \otimes \sigma_x \otimes \openone \end{array} \right)  & {}\ \ \ & \mathbf{q}_4 = \left( \begin{array}{c} \openone \otimes \openone \otimes \openone \otimes \sigma_x\\  \openone \otimes \openone \otimes \sigma_x \otimes \sigma_y \\ \openone \otimes \openone \otimes \sigma_x \otimes \sigma_z \end{array} \right) \end{eqnarray*}

\noindent We now rotate qubits 2 and 3 to the Bell basis by performing a Bell gate, giving
\begin{eqnarray*} \mathbf{q}_1 = \left( \begin{array}{c} \sigma_z \otimes \sigma_x \otimes \openone \otimes \openone\\  -\sigma_y \otimes \sigma_x \otimes \openone \otimes \openone \\ \sigma_x \otimes \openone \otimes \openone \otimes \openone \end{array} \right) & {}\ \ \ & \mathbf{q}_2 = \left( \begin{array}{c} \openone \otimes \sigma_x \otimes \openone \otimes \openone\\  - \sigma_x \otimes \sigma_y \otimes \sigma_x \otimes \openone \\ - \sigma_x \otimes \sigma_z \otimes \sigma_x \otimes \openone \end{array} \right) \\
\mathbf{q}_3 = \left( \begin{array}{c} \openone \otimes \openone \otimes \sigma_x \otimes \openone\\  - \openone \otimes \sigma_x \otimes \sigma_y \otimes \sigma_x \\ \openone \otimes \sigma_x \otimes \sigma_z \otimes \sigma_x \end{array} \right)  & {}\ \ \ & \mathbf{q}_4 = \left( \begin{array}{c} \openone \otimes \openone \otimes \openone \otimes \sigma_x\\  \openone \otimes \openone \otimes \sigma_x \otimes \sigma_y \\ \openone \otimes \openone \otimes \sigma_x \otimes \sigma_z \end{array} \right) \end{eqnarray*}

Now we perform the measurement on the pair (2,3) in the Bell basis by introducing two further qubits and performing two CNOT operations between (3,5) and (2,6), leaving us with the complete set of descriptors
\begin{eqnarray*} & \mathbf{q}_1 = \left( \begin{array}{c} \sigma_z \otimes \sigma_x \otimes \openone \otimes \openone \otimes \openone \otimes \openone \\  -\sigma_y \otimes \sigma_x \otimes \openone \otimes \openone \otimes \openone \otimes \openone \\ \sigma_x \otimes \openone \otimes \openone \otimes \openone \otimes \openone \otimes \openone \end{array} \right) & \\ & \mathbf{q}_4 = \left( \begin{array}{c} \openone \otimes \openone \otimes \openone \otimes \sigma_x \otimes \openone \otimes \openone \\  \openone \otimes \openone \otimes \sigma_x \otimes \sigma_y \otimes \openone \otimes \openone \\ \openone \otimes \openone \otimes \sigma_x \otimes \sigma_z \otimes \openone \otimes \openone \end{array} \right) & \\ & \mathbf{q}_5 = \left( \begin{array}{c} \openone \otimes \openone \otimes \openone \otimes \openone \otimes \sigma_x \otimes \openone \\ \openone \otimes \sigma_x \otimes \sigma_z \otimes \sigma_x \otimes \sigma_y \otimes \openone \\ \openone \otimes \sigma_x \otimes \sigma_z \otimes \sigma_x \otimes \sigma_z \otimes \openone  \end{array} \right) & \\ & \mathbf{q}_6 = \left( \begin{array}{c} \openone \otimes \openone \otimes \openone \otimes \openone \otimes \openone \otimes \sigma_x \\ -\sigma_x \otimes \sigma_y \otimes \sigma_x \otimes \openone \otimes \openone \otimes \sigma_y  \\ -\sigma_x \otimes \sigma_z \otimes \sigma_x \otimes \openone \otimes \openone \otimes \sigma_z \end{array} \right) & \\ & \mathbf{q}_2 = \left( \begin{array}{c} \openone \otimes \sigma_x \otimes \openone \otimes \openone \otimes \openone \otimes \sigma_x \\  - \sigma_x \otimes \sigma_y \otimes \sigma_x \otimes \openone \otimes \openone \otimes \sigma_x \\ - \sigma_x \otimes \sigma_z \otimes \sigma_x \otimes \openone \openone \otimes \openone \end{array} \right) & \\ & \mathbf{q}_3 = \left( \begin{array}{c} \openone \otimes \openone \otimes \sigma_x \otimes \openone \otimes \sigma_x \otimes \openone \\  - \openone \otimes \sigma_x \otimes \sigma_y \otimes \sigma_x \otimes \sigma_x \otimes \openone \\ \openone \otimes \sigma_x \otimes \sigma_z \otimes \sigma_x  \otimes \openone \otimes \openone\end{array} \right) & \end{eqnarray*}

If we now look at these descriptors, we can see some very interesting things. First, let us look at the density matrices for the pairs (2,3) and (1,4). For both pairs, $\rho_{ab}=\openone \otimes \openone$ and $\rho_{a,b}=\openone$, from which we can see that the pairs are not entangled. Next, we note that neither pair is in a pure state, as we cannot reduce their descriptors simply to the elements on the Hilbert spaces for those systems. Furthermore, if we look at the pairs (1,2) and (3,4) which were originally entangled, we see that they too have $\rho_{ab}=\openone\otimes\openone, \ \rho_{a,b}=\openone$, so are no longer entangled. The pairs that \emph{are} entangled at this point at (3,5) and (2,6), which have become so through the measurement interaction.

The exact forms of these descriptors is also interesting. We know that, because of the locality of interactions in the Deutsch-Hayden representation, the only way in which a descriptor can have a non-trivial dependence on \h{a} is by interacting either directly with system $a$, or with something that has interacted with it. We can therefore trace, not only dependencies on specific quantities as Deutsch and Hayden did, but also trace non-trivial elements of a descriptor on a certain Hilbert space associated with another system. If we look at the descriptors for qubits 2 and 3 we can see that they have dependencies on many spaces, and we can trace where they have come from. Qubit 2 has non-trivial elements on \h{1,2,3,6}. The dependence on \h{1} comes from the original Bell state that the pair (1,2) shared. The \h{3} element comes from the Bell gate interaction with qubit 3, and the \h{6} element from the CNOT operation with qubit 6. Similarly we can traces the dependencies of qubit 3, with are on \h{2,3,4,5}. We can also see that qubits 1 and 4 remain as they were at the beginning: they have not interacted any further in the protocol, and this is reflected in the components of their descriptors. Finally we note that the Hilbert space dependencies of qubits 5 and 6 are the same as those of the qubits that they interacted with in the CNOT operations.

In order to fully `swap' the entanglement, we are going to need to look at the various descriptors relative to the four binary numbers stored in qubits 5 and 6. By construction, the pair (2,3) will be in one of the four Bell states, but what about the pair (1,4)? First, we need the four-system version of our expression (\ref{jfda}) for relative descriptors. This is easily shown to be
$$ q^\prime_{12ij} = q_{a12ij} (1 + \sum_{mn} \av{q^\prime_{34nm}} q_{34nm} )$$

Now we use the fact that
$$ \av{q^\prime_{34nm}} = \av{\sigma_n\otimes \sigma_m}_{\ketbra{\beta}{\beta}}$$

\noindent where $\{\ketbra{\beta}{\beta}\}$ is the set of states of (3,4) relative to which we are finding the descriptors for (1,2). If $\{\ketbra{\beta}{\beta}\} = \{ \ketbra{00}{00}, \ketbra{01}{01}, \ketbra{10}{10}, \ketbra{11}{11}\}$ then we have
\begin{eqnarray*} q^\prime_{12ij} = q_{12ij} ( 1 \pm (q_{3z} + q_{4z}) + q_{3z}q_{4z}) & {} \ \ \ {} & \frac{00}{11}\\
q^\prime_{12ij} = q_{12ij} ( 1 \pm (q_{3z} - q_{4z}) - q_{3z}q_{4z}) & {} \ \ \ {} & \frac{01}{10}
\end{eqnarray*}

If we look at the descriptors for qubit 1 and 4, and the $\sigma_z$-components of qubits 5 and 6, we see that the only nonzero components of the relative descriptors are
$$ q_{1x}q_{4x}q_{5z}, \ \ q_{1y}q_{4y}q_{5z}q_{6z}, \ \ q_{1z}q_{4z}q_{6z}$$

\noindent with their appropriate signs. We can therefore represent the relative descriptors in this case by
\begin{eqnarray*} q_{1x} &\longrightarrow & (++--) q_{1x}q_{5z}\\
q_{4z} &\longrightarrow & (+-+-) q_{4z}q_{6z}\end{eqnarray*}

\noindent which gives us the full set of descriptors relative to the set of measurements on (5,6) $\{00,01,10,11\}$:
\begin{eqnarray*} & \mathbf{q}^\prime_1 = (++--) \left( \begin{array}{c} \sigma_z \otimes \openone \otimes \sigma_z \otimes \sigma_x \otimes \sigma_z \otimes \openone \\  -\sigma_y \otimes \openone \otimes \sigma_z \otimes \sigma_x \otimes \sigma_z \otimes \openone \\ \sigma_x \otimes \openone \otimes \openone \otimes \openone \otimes \openone \otimes \openone \end{array} \right) & \\ & \mathbf{q}^\prime_4 = (+-+-) \left( \begin{array}{c} \openone \otimes \openone \otimes \openone \otimes \sigma_x \otimes \openone \otimes \openone \\  - \sigma_x \otimes \sigma_z \otimes \openone \otimes \sigma_y \otimes \sigma_z \otimes \openone \\ - \sigma_x \otimes \sigma_z \otimes \openone \otimes \sigma_z \otimes \openone \otimes \sigma_z \end{array} \right) & \end{eqnarray*}

If we now look at the reduced form of these descriptors on \h{1,4} only, we see that the simply reduced forms retain all the same average values. We may therefore write the descriptors as
$$ \mathbf{q}^\prime_1 = (++--) \left( \begin{array}{c} \sigma_z \otimes \sigma_x   \\  -\sigma_y \otimes  \sigma_x  \\ \sigma_x \otimes \openone \end{array} \right) {}\ \ \   \mathbf{q}^\prime_4 = (+-+-) \left( \begin{array}{c} \openone \otimes \sigma_x  \\  - \sigma_x  \otimes \sigma_y \\ - \sigma_x  \sigma_z \end{array} \right)$$

\noindent which are the descriptors corresponding to the four (pure) Bell states.

The first thing to note when analysing the flow of dependencies is that finding the relative descriptors requires the components $q_{5z}$ and $q_{6z}$. Therefore, the `swapping' of entanglement can only happen when these components are transmitted to qubits 1 and 4. That is, each qubit needs two bits of classical communication sent to it before the two qubits can become entangled.

We therefore have the transmission of the dependencies of $q_{5z,6z}$ to qubits 1 and 4, and it is only once this has happened that we see the entanglement being swapped. Qubits 5 and 6 have dependencies on all of qubits 1 to 4, and these are transmitted to qubits 1 and 4, possibly separately. The dependencies on (2,3) work out to be the correlation between states of (2,3) and (1,4), and those on (1,4) give rise to the entanglement between qubits 1 and 4.

Thus we see that there is nothing non-local about entanglement swapping. There is no instantaneous ``transmission" of entanglement to the qubit pair (1,4) because of the measurement on the pair (2,3), and no superluminal signalling between qubit 1 and 4 to tell them that they are now entangled with each other. The entanglement dependencies (and those which give the correlation between the two pairs - a facet of entanglement swapping that is often overlooked) are transmitted to qubits 1 and 4 separately through qubits 5 and 6. As it is only their $\sigma_z$ components which are used, we can describe them as two bits of data being transmitted over a classical communication channel, giving the result of the measurement of the pair (2,3) in the Bell basis. This, then, is a completely local description of entanglement swapping.

\section{Conclusions}

In this paper we have developed the Deutsch-Hayden formalism to cover areas that it could not originally, specifically dealing with measurement and mixed states. We have also seen that, when considering the picture of information flow given to us by the descriptors, we can be confident that this is a unique picture, not dependent on the exact form chosen for those descriptors. Finally we have considered the entanglement swapping protocol, and seen that the usual description of it as creating entanglement non-locally is incorrect, and that in fact the entanglement is transmitted entirely locally. We have therefore seen how the Deutsch-Hayden approach is transparent to the notions of locality and unitarity in quantum mechanics, in a way that the standard formalism is not.


\begin{thebibliography}{99}

\bibitem{dh} D. Deutsch and P. Hayden, Proc. R. Soc. Lond. A 456 1759-1774 (2000).

\bibitem{gottesman} D. Gottesman, quant-ph/9807006 (1998).

\bibitem{fano} U. Fano, Reviews of Modern Physics 29(1) (1957).

\bibitem{nandc} M. Nielsen and I. Chuang, \emph{Quantum Computation and Quantum Information} (CUP, Cambridge, 2000).

\bibitem{handj} R. Horn and C. Johnson, \emph{Matrix Analysis} (CUP, Cambridge, 1985).

\bibitem{entswap} M. Zukowski, A. Zeilinger, M. Horne and A. Ekert, Phys. Rev. Lett. 71 (1993)

\end{thebibliography}
\end{document}